\documentclass[prd,showpacs]{revtex4}
\usepackage{amsmath,amssymb}
\usepackage{xcolor}

\begin{document}
\title{Finite temperature Casimir effect of a Lorentz-violating scalar with higher order derivatives}
\author{Andrea Erdas}
\email{aerdas@loyola.edu}
\affiliation{Department of Physics, Loyola University Maryland, 4501 North Charles Street,
Baltimore, Maryland 21210, USA}
\begin {abstract} 
In this work I study the finite temperature Casimir effect caused by a complex and massive scalar field that breaks Lorentz invariance in a CPT-even, aether-like manner. 
The Lorentz invariance breaking is caused by a constant space-like vector directly coupled to higher order field derivatives. This vector needs to be space-like in order to avoid non-causality problems that will arise with a time-like vector. I investigate the two scenarios of the scalar field satisfying either Dirichlet or mixed boundary conditions on a pair of flat parallel plates. I use the generalized zeta function technique that enables me to obtain the Helmholtz free energy and the Casimir pressure when the Casimir plates are in thermal equilibrium with a heat reservoir at finite temperature. I investigate two different directions of the unit vector, parallel and perpendicular to the plates. I examine both scenarios for both types of boundary conditions and, in both cases and for the two different boundary conditions, I obtain simple analytic expressions of the Casimir energy and pressure in the three asymptotic limits of small plate distance, high temperature, and large mass, examining all combinations of boundary conditions, unit vector direction, and asymptotic limits.

\end {abstract}
\pacs{03.70.+k, 11.10.-z, 11.30.Cp.}
\maketitle
\section{Introduction}
\label{1}
Casimir's theoretical prediction was made 75 years ago \cite{Casimir:1948dh}. 
In his initial understanding, two uncharged and perfectly conducting plates facing each other in vacuum will attract, due to quantum field theory effects.
The initial experimental verification came from Sparnaay \cite{Sparnaay:1958wg} ten years after Casimir published his work, and was loosely consistent with his theoretical prediction. 
The effect has since been confirmed by many experiments of increasing accuracy \cite{Bordag:2001qi,Bordag:2009zz} and is now on very solid grounds. 
Casimir's  original paper studied quantum fluctuations of the photon field in a vacuum, and discovered these fluctuations produce an attraction between the plates. Some time later, it was understood that
quantum fluctuations of other fields produce Casimir forces too, which depend strongly on the boundary conditions and shape of the plates \cite{Boyer:1974,Boyer:1968uf}.
Dirichlet or Neumann boundary conditions cause attraction of the plates, mixed (Dirichlet-Neumann) boundary conditions cause repulsion. 

The breaking of Lorentz symmetry has been vigorously investigated by high energy theorists in the last twenty to thirty years, with many models being proposed that cause a space-time anisotropy  \cite{Ferrari:2010dj,Ulion:2015kjx,Alfaro:1999wd,Alfaro:2001rb} within the larger context of seeking a theory of quantum gravity. Some models propose variation of some coupling constants \cite{Kostelecky:2002ca,Anchordoqui:2003ij,Bertolami:1997iy} and, in string theory \cite{Kostelecky:1988zi}, some vector and tensor field components can acquire non-vanishing vacuum expectation values that generate a spontaneous Lorentz symmetry breaking  at the Planck energy scale. A detailed list of papers that study various consequences of Lorentz symmetry breaking is available in Refs. \cite{Cruz:2017kfo,ADantas:2023wxd}. 
All these models break Lorentz invariance at the Planck scale, with repercussions on the space-time anisotropy that are observable at much lower energy through the Casimir effect. Both scalar fields \cite{Cruz:2017kfo,Cruz:2018bqt} and fermion fields \cite{daSilva:2019iwn} that break Lorentz symmetry have been investigated within the context of modifications to the standard Casimir effect. 
Consequences of the existence of Lorentz violation in the Casimir effect have also been studied in the case of Lorentz-breaking extensions of QED \cite{Frank:2006ww,Kharlanov:2009pv,Martin-Ruiz:2016lyy}. In the last few years, some authors have examined the case of a real scalar field  in vacuum \cite{Cruz:2017kfo,ADantas:2023wxd,Cruz:2020zkc} and in a medium at finite temperature \cite{Cruz:2018bqt}, or a complex scalar field in vacuum with a magnetic field \cite{Erdas:2020ilo}, and in a medium at finite temperature in the presence of a magnetic field \cite{Erdas:2021xvv}. 
These more recent papers examine a modified Klein-Gordon model that breaks Lorentz symmetry in a CPT-even, aether-like manner. While the majority of these papers investigate a situation where  Lorentz violation is due to the presence of a constant vector directly coupled to the first derivatives of the scalar field, one of them \cite{ADantas:2023wxd}, examines the scenario where a constant space-like vector is directly coupled to higher order derivatives of the field.

Several authors studied the finite temperature Casimir effect in Lorentz symmetric spacetime \cite{Fierz:1960zq,Mehra:1967wf,CougoPinto:1998jg,Erdas:2013jga,Erdas:2013dha,Aleixo:2021cfy}, or in spacetime where Lorentz violation is implemented by a constant vector coupled to the first order derivatives of the field \cite{Cruz:2018bqt,Erdas:2020ilo,Erdas:2021xvv}, but there has not been a study of the finite temperature Casimir effect of a charged scalar field that breaks the Lorentz symmetry in a CPT-even, aether-like manner implemented by a unit space-like vector coupled to arbitrarily high order derivatives of the field. This paper intends to fill that gap and provide theoretical predictions of the finite temperature effects on the quantum vacuum of the modified Klein-Gordon model presented in Ref. \cite{ADantas:2023wxd}.
In this paper I will first obtain the free energy, then the Casimir pressure of a Lorentz-violating scalar field by studying a model similar to the one first presented in Ref.  \cite{ADantas:2023wxd}: a charged scalar field that breaks Lorentz symmetry and satisfies either Dirichlet or mixed boundary conditions on a pair of large parallel plates. I will not examine the case where the field obeys Neumann boundary conditions on the plates, because it produces the same results obtained with Dirichlet boundary conditions.

In Sec. \ref{2} of this work I introduce the theoretical model of a complex scalar field that breaks Lorentz symmetry in an aether-like and CPT-even manner, by way of the coupling of a space-like unit vector to higher derivatives of the field, and use the zeta function technique \cite{Hawking:1976ja,Elizalde:1988rh,Elizalde:2007du} to obtain an expression of the model's zeta function containing integrals and infinite sums. I obtain the zeta function for the case where the unit vector is parallel to the plates and perpendicular to the plates. In Sec. \ref{3} I examine the case of a Lorentz asymmetry in the direction parallel to the Casimir plates and calculate the free energy, obtaining simple analytic expressions for the free energy in the short plate distance limit, high temperature limit, and large mass limit. In Sec. \ref{4} I investigate the case of Lorentz anisotropy perpendicular to the plates, calculate the free energy, and obtain simple expressions for it in the three limits listed above. In Sec. \ref{5} I calculate the Casimir pressure for all the cases described above. In all these sections I focus first on the scalar field obeying Dirichlet boundary conditions at the plates, and then on it obeying mixed boundary conditions at the plates. My conclusions and a detailed discussion of my results are in Sec. \ref{6}.
\section{The model and its zeta function}
\label{2}

In this paper I use $\hbar=c=1$ and study the Casimir effect due to a complex scalar field $\phi$ of mass $M$ that breaks Lorentz symmetry in an aether-like and CPT-even  manner. The Lorentz-symmetry breaking is implemented by a constant space-like unit vector $u^\mu$ coupled to higher order derivatives of $\phi$, as in the theoretical model introduced by Ref. \cite{ADantas:2023wxd}. The modified Klein Gordon equation for $\phi$ is
\begin{equation}
[\Box 
+l^{2(\epsilon -1)}(u\cdot\partial)^{2\epsilon}+M^2]\phi=0,
\label{KG}
\end{equation}
where $u^\mu$ points in the direction in which the Lorentz symmetry is broken, the length $l$ is of order of the inverse of the energy scale at which the Lorentz symmetry is broken, and  $\epsilon$ is a positive integer parameter that plays the same role as the critical exponent does for fermion fields in Horava-Lifshitz theories \cite{daSilva:2019iwn,Farias:2011aa,Erdas:2025etj}. The case with $\epsilon=1$ has already been studied in Ref. \cite{Cruz:2017kfo} and, for the magnetic case, in \cite{Erdas:2020ilo}. In this work, I examine $\epsilon\ge 2$ and, to avoid non-causality problems, I take $u^\mu$ to be spacelike only. 

My aim is to study how this type of space-time anisotropy modifies the Casimir effect at finite temperature. I will assume the Casimir plates to be in thermal equilibrium with a heat reservoir at temperature $T$ and will calculate thermal effects on their Casimir energy caused by the Lorentz-violating scalar field.  I consider two square plates of side $L$ perpendicular to the $z$ axis, located at $z=0$ and $z=a$. 
I will use the generalized zeta function technique \cite{Hawking:1976ja,Elizalde:1988rh,Elizalde:2007du} to study this problem, and will investigate Dirichlet and mixed boundary conditions of the field $\phi$ at the plates. Investigating Neumann boundary conditions is trivial, since it produces the same results found with Dirichlet boundary conditions. I will examine the two cases of $u^\mu$ parallel to the plates, and $u^\mu$ perpendicular to the plates.
For this system the imaginary time formalism of finite temperature field theory is convenient, and it allows only scalar field configurations satisfying the following periodic condition:
\begin{equation}
\phi(x,y,z,\tau)=\phi(x,y,z,\tau+\beta)
\label{periodic}
\end{equation}
for any value of the Euclidean time $\tau$, where $\beta = T^{-1}$ is the periodic length in the Euclidean time axis. The Helmholtz free energy of the field-plates system is
\begin{equation}
F=\beta^{-1}\log \,\det \left(D_{\rm E}|{\cal F}_a\right),
\label{F_1}
\end{equation}
where $D_{\rm E}$ is the "modified" Klein-Gordon operator in Euclidean time 
\begin{equation}
D_{\rm E}=-\partial^2_\tau-\nabla^{2}+M^2+\ell^{2(\epsilon-1)}(u\cdot\partial)^{2\epsilon}
\label{D_E}
\end{equation}
and the symbol ${\cal F}_a$ indicates the set of eigenfunctions of $D_{\rm E}$ which satisfy the periodic condition of Eq. (\ref{periodic}), and Dirichlet or mixed boundary conditions at the plates. According to the generalized zeta function technique, the free energy is related to the derivative of the zeta function $\zeta(s)$ of the operator $D_{\rm E}$ by the following
\begin{equation}
F=-\beta^{-1}\zeta'(0).
\label{F_2}
\end{equation}
To construct this zeta function I need the eigenvalues of the operator $D_{\rm E}$ that satisfy the boundary conditions specified above. The eigenvalues 
of $-\partial^2_\tau-\nabla^{2}+M^2$ are
\begin{equation}
k^2_\tau+k^2_x+k^2_y+k^2_z+M^2,
\label{eigenvalues_1}
\end{equation}   
where
\begin{equation}
k_\tau= {2\pi m\over \beta},
\label{periodic_1}
\end{equation}   
with $m=0, \pm 1, \pm 2, \cdots$, $k_x$ and $k_y$ can take any real value, while 
\begin{equation}
k_z={n\pi\over a},
\label{kz_D}
\end{equation}
with $n=1,2,3,\cdots$, for Dirichlet boundary conditions, and 
\begin{equation}
k_z=\left(n+\frac{1}{2}\right)\frac{\pi}{ a},
\label{kz_mixed}
\end{equation}
with $n=0,1,2,3,\cdots$, for mixed boundary conditions.
The eigenvalues of $\ell^{2(\epsilon-1)}(u\cdot\partial)^{2\epsilon}$ depend on the direction of $u^\mu$. When $u^\mu$ is parallel to the plates
\begin{equation}
u^\mu=\left(0,{1\over \sqrt{2}},{1\over \sqrt{2}},0\right),
\label{upar}
\end{equation}
the eigenvalues of $\ell^{2(\epsilon-1)}(u\cdot\partial)^{2\epsilon}$ are
\begin{equation}
l^{2(\epsilon - 1)}(-1)^\epsilon\left({k^{2\epsilon}_x+k^{2\epsilon}_y\over2}\right),
\label{eigenvalues_2}
\end{equation}
where $k_x$ and $k_y$ are real. When $u^\mu$ is perpendicular to the plates
\begin{equation}
u^\mu=\left(0,0,0,1\right),
\label{uperp}
\end{equation}
the eigenvalues of $\ell^{2(\epsilon-1)}(u\cdot\partial)^{2\epsilon}$ are
\begin{equation}
l^{2(\epsilon - 1)}(-1)^\epsilon k^{2\epsilon}_z,
\label{eigenvalues_3}
\end{equation}
where $k_z$ takes the discrete values shown in Eqs. (\ref{kz_D}) or (\ref{kz_mixed}). 

Adding the eigenvalues shown in Eqs. (\ref{eigenvalues_1}) and (\ref{eigenvalues_2}), I obtain the zeta function for $u^\mu$ parallel to the plates
\begin{equation}
\zeta(s)=\mu^{2s}{L^2\over (2\pi)^2}\sum_{m,n}\int dk_x dk_y \left[k^2_\tau+k^2_x+k^2_y+k^2_z+M^2+l^{2(\epsilon - 1)}(-1)^\epsilon\left({k^{2\epsilon}_x+k^{2\epsilon}_y\over2}\right)\right]^{-s},
\label{zeta_1}
\end{equation}
where $k_\tau$ and $k_z$ take discrete values, as shown by Eqs. (\ref{periodic_1}), (\ref{kz_D}), and (\ref{kz_mixed}), and the sums are over these discrete values. Notice that, as it is done routinely when applying the zeta function technique \cite{Hawking:1976ja}, I use the parameter $\mu$ with dimension of mass to keep $\zeta(s)$ dimensionless for all values of $s$. Next, I do a change of integration variables from cartesian to plane polar coordinates, $(k_x, k_y) \rightarrow (k, \theta)$, and obtain
\begin{equation}
\zeta(s)=\mu^{2s}{L^2\over (2\pi)^2}\sum_{m,n}\int_0^\infty k dk \int_0^{2\pi} d\theta \left[k^2_\tau+k^2+k^2_z+M^2+l^{2(\epsilon - 1)}(-1)^\epsilon k^{2\epsilon}\left({\cos^{2\epsilon}\theta+\sin^{2\epsilon}\theta\over 2}
\right)\right]^{-s},
\label{zeta_2}
\end{equation}
then, using the identity
 \begin{equation}
z^{-s}=\frac{1}{ \Gamma(s)}\int_0^\infty dt\, t^{s-1}e^{-zt},
\label{z-s}
\end{equation}
 where $\Gamma(s)$ is the Euler gamma function, I write the zeta function as
\begin{equation}
\zeta(s)={\mu^{2s}\over \Gamma(s)}{L^2\over (2\pi)^2}\sum_{m,n}\int_0^\infty k dk \int_0^{2\pi} d\theta\int_0^\infty dt\, t^{s-1}e^{-(k^2_\tau+k^2+k^2_z+M^2) t} e^{- l^{2(\epsilon - 1)}(-1)^\epsilon k^{2\epsilon}\left({\cos^{2\epsilon}\theta+\sin^{2\epsilon}\theta\over 2}
\right)t}.
\label{zeta_3}
\end{equation}
Since the length $l \ll a$, I expand up to first order in this small length and find
\begin{equation}
\zeta(s)=\zeta_R(s)+{\tilde \zeta}(s),
\label{zeta_4}
\end{equation}
where $\zeta_R(s)$ is the "regular" zeta function in the absence of Lorentz violation
\begin{equation}
\zeta_R(s)={\mu^{2s}\over \Gamma(s)}{L^2\over (2\pi)^2}\sum_{m,n}\int_0^\infty k dk \int_0^{2\pi} d\theta\int_0^\infty dt\, t^{s-1}e^{-(k^2_\tau+k^2+k^2_z+M^2) t} ,
\label{zeta_5}
\end{equation}
and ${\tilde \zeta}(s)$ contains the Lorentz violating part 
\begin{equation}
{\tilde \zeta}(s)=-{\mu^{2s}\over \Gamma(s)}{L^2\over (2\pi)^2}l^{2(\epsilon - 1)}(-1)^\epsilon \sum_{m,n}\int_0^\infty k^{2\epsilon+1} dk \int_0^{2\pi}  \left({\cos^{2\epsilon}\theta+\sin^{2\epsilon}\theta\over 2}
\right)d\theta\int_0^\infty dt\, t^{s}e^{-(k^2_\tau+k^2+k^2_z+M^2) t}.
\label{zeta_6}
\end{equation}
I will focus only on ${\tilde \zeta}(s)$, since $\zeta_R(s)$ and the corresponding free energy have been evaluated in many papers in the past. After the $k$ and $\theta$ integrations are done, I obtain
\begin{equation}
{\tilde \zeta}(s)=-{\mu^{2s}\over \Gamma(s)}{L^2\over 4\pi}l^{2(\epsilon - 1)}(-1)^\epsilon {(2\epsilon - 1)!!\over 2^\epsilon}\sum_{m,n}
\int_0^\infty dt\, t^{s-\epsilon -1}e^{-(k^2_\tau+k^2_z+M^2) t}.
\label{zeta_7}
\end{equation}
I will use ${\tilde \zeta}(s)$ of Eq. (\ref{zeta_7}) in the next sections of this work to obtain the Lorentz violating corrections to the free energy and Casimir pressure in the case of $u^\mu$ parallel to the plates.

Combining the eigenvalues shown in Eqs. (\ref{eigenvalues_1}) and (\ref{eigenvalues_3}), I obtain the zeta function for $u^\mu$ perpendicular to the plates
\begin{equation}
\zeta(s)=\mu^{2s}{L^2\over (2\pi)^2}\sum_{m,n}\int dk_x dk_y \left[k^2_\tau+k^2_x+k^2_y+k^2_z+M^2+l^{2(\epsilon - 1)}(-1)^\epsilon k^{2\epsilon}_z\right]^{-s},
\label{zeta_8}
\end{equation}
then I do a change of integration variables from cartesian to plane polar coordinates, $(k_x, k_y) \rightarrow (k, \theta)$, use the identity of Eq. (\ref{z-s}), and find
\begin{equation}
\zeta(s)={\mu^{2s}\over \Gamma(s)}{L^2\over (2\pi)^2}\sum_{m,n}\int_0^\infty k dk \int_0^{2\pi} d\theta\int_0^\infty dt\, t^{s-1}e^{-(k^2_\tau+k^2+k^2_z+M^2) t} e^{- l^{2(\epsilon - 1)}(-1)^\epsilon k^{2\epsilon}_zt}.
\label{zeta_9}
\end{equation}
I do the straightforward $k$ and $\theta$ integrations, then expand expand up to first order in the small length $l$ and find $\zeta(s)=\zeta_R(s)+{\tilde \zeta}(s)$,
where $\zeta_R(s)$ is given by Eq. (\ref{zeta_5}), and
\begin{equation}
{\tilde \zeta}(s)=-{\mu^{2s}\over \Gamma(s)}{L^2\over 4\pi}l^{2(\epsilon - 1)}(-1)^\epsilon \sum_{m,n}k_z^{2\epsilon}
\int_0^\infty dt\, t^{s -1}e^{-(k^2_\tau+k^2_z+M^2) t},
\label{zeta_10}
\end{equation}
is the Lorentz violating part of the zeta function. Again, I will use ${\tilde \zeta}(s)$ of Eq. (\ref{zeta_10}) in the next sections of this paper to obtain the Lorentz violating corrections to the free energy and pressure in the case of $u^\mu$ perpendicular to the plates.
\section{Free energy for $u^\mu$ parallel to the plates}
\label{3}
In this section I will start from Eq. (\ref{zeta_7}) and use that zeta function to obtain the Lorentz violating corrections to the free energy for the case of $u^\mu$ parallel to the plates. Eq. (\ref{zeta_7}) gives the exact form of the Lorentz violating correction to the zeta function when $u^\mu$ is parallel to the plates but it can only be reduced to a simple analytic form in three asymptotic cases: $a^{-1}\gg T , m$ (small plate distance); $m\gg T , a^{-1}$ (large mass); and $T \gg m, a^{-1}$ (high temperature). I will examine each of these three cases, starting with the small plate distance limit and will obtain the free energy corrections for Dirichlet and mixed boundary conditions.

In the small plate distance limit, I do a Poisson resummation of the $m$-sum in the zeta function
\begin{equation}
\sum_{m=-\infty}^\infty e^{-k^2_\tau t}={\beta\over \sqrt{\pi t}}\left({1\over 2}+\sum_{m=1}^\infty e^{-\beta^2m^2/4t}\right),
\label{Poisson_1}
\end{equation}
use it into Eq. (\ref{zeta_7}) and obtain
\begin{equation}
{\tilde \zeta}(s)={\tilde \zeta}_0(s)+{\tilde \zeta}_T(s),
\label{zeta_11}
\end{equation}
where ${\tilde \zeta}_0(s)$ is the Lorentz violating correction to the vacuum zeta function
\begin{equation}
{\tilde \zeta}_0(s)=-{\mu^{2s}\over \Gamma(s)}{L^2\beta\over 8\pi^{3/2}}l^{2(\epsilon - 1)}(-1)^\epsilon {(2\epsilon - 1)!!\over 2^\epsilon}\sum_{n}
\int_0^\infty dt\, t^{s-\epsilon -3/2}e^{-(k^2_z +M^2)t},
\label{zeta_12}
\end{equation}
and ${\tilde \zeta}_T(s)$ is the Lorentz violating thermal correction to the zeta function
\begin{equation}
{\tilde \zeta}_T(s)=-{\mu^{2s}\over \Gamma(s)}{L^2\beta\over 4\pi^{3/2}}l^{2(\epsilon - 1)}(-1)^\epsilon {(2\epsilon - 1)!!\over 2^\epsilon}\sum_{n}
\int_0^\infty dt\, t^{s-\epsilon -3/2}e^{-(k^2_z +M^2)t}\sum_{m=1}^\infty e^{-\beta^2m^2/4t}.
\label{zeta_13}
\end{equation}
Since $M\ll a^{-1}$, I take
\begin{equation}
e^{-M^2t}\simeq 1-M^2t,
\label{mass_1}
\end{equation}
then do the $t$ integration, to obtain
\begin{equation}
{\tilde \zeta}_0(s)=-{\mu^{2s}\over \Gamma(s)}{L^2\beta\over 8\pi^{3/2}}l^{2(\epsilon - 1)}(-1)^\epsilon {(2\epsilon - 1)!!\over 2^\epsilon}
\left[\Gamma(s-\epsilon - {1\over 2})\sum_{n}k_z^{2\epsilon+1-2s}-M^2\Gamma(s-\epsilon + {1\over 2})\sum_{n}k_z^{2\epsilon-1-2s}
\right].
\label{zeta_14}
\end{equation}
In the case of Dirichlet boundary conditions, I use Eq. (\ref{kz_D}) for $k_z$ and evaluate the infinite sums
\begin{equation}
\sum_{n=1}k_z^{2x}=\left({\pi\over a}\right)^{2x}\zeta_R(-2x),
\label{zeta_15}
\end{equation}
where $\zeta_R(z)$ is the Riemann zeta function, while for mixed boundary conditions I use Eq. (\ref{kz_mixed}) for $k_z$ and evaluate the infinite sums 
\begin{equation}
\sum_{n=0}k_z^{2x}=\left({\pi\over a}\right)^{2x}\zeta_H(-2x,{1\over 2}),
\label{zeta_16}
\end{equation}
in terms of the Hurwitz zeta function $\zeta_H(z,s)=\sum\limits_{\substack{n =0}}^\infty (n+s)^{-z}$. I use Eq. (\ref{zeta_15}) and, for Dirichlet boundary conditions, find
\begin{eqnarray}
{\tilde \zeta}_0(s)&=&-{\mu^{2s}\over \Gamma(s)}{L^2\beta\over 8\pi^{3/2}}l^{2(\epsilon - 1)}(-1)^\epsilon {(2\epsilon - 1)!!\over 2^\epsilon}\left({\pi\over a }\right)^{2\epsilon+1-2s}
\left[\Gamma(s-\epsilon - {1\over 2})\zeta_R({2s-2\epsilon-1})\right.
\nonumber \\
&-&\left.{M^2a^2\over\pi^2}\Gamma(s-\epsilon + {1\over 2})\zeta_R({2s-2\epsilon+1})
\right].
\label{zeta_17}
\end{eqnarray}
The free energy is related to the derivative of the zeta function by $F=-\beta^{-1}\zeta'(0)$, so I need to evaluate ${\tilde \zeta}_0(s)$ near $s=0$.
I find that, for $s\ll 1$
\begin{equation}
{f(s)\over \Gamma (s)}\simeq f(0)s +{\cal O}(s^2)
\label{gamma_2}.
\end{equation}
I also use the following
\begin{equation}
\Gamma (-n+{1\over 2})=(-1)^n{2^n\sqrt{\pi}\over (2n-1)!!}
\label{gamma_3},
\end{equation}
where $n$ is a non-negative integer and, for Dirichlet boundary condition, I obtain 
\begin{equation}
{\tilde F}_0=-{L^2\over 4a^3}\left( {l\over a}\right)^{2(\epsilon - 1)}{\pi }^{2\epsilon}
\left[{\zeta_R({-2\epsilon-1})\over 2\epsilon +1}+{M^2a^2\over2\pi^2}\zeta_R({-2\epsilon+1})
\right].
\label{F_3}
\end{equation}
where ${\tilde F}_0$ is the Lorentz violating correction to the vacuum Casimir energy. This result fully agrees with the result of Ref. \cite{ADantas:2023wxd} which is obtained using a different method, the Abel-Plana method. 

To obtain the Lorentz violating correction to the vacuum Casimir energy in the case of mixed boundary conditions, all I need to do, according to Eqs. (\ref{zeta_15}) and (\ref{zeta_16}), is replace $\zeta_R(x)$ with $\zeta_H(x,{1\over 2})$. Once I do that and use 
\begin{equation}
\zeta_H(x,{1\over 2})=(2^x-1)\zeta_R(x),
\label{zh_1}
\end{equation}
I find that, for mixed boundary conditions, the Lorentz violating correction to the vacuum Casimir energy is
\begin{equation}
{\tilde F}_0={L^2\over 4a^3}\left( {l\over a}\right)^{2(\epsilon - 1)}{\pi }^{2\epsilon}
\left[{(1-2^{-2\epsilon-1})\over 2\epsilon +1}\zeta_R({-2\epsilon-1})+{M^2a^2\over2\pi^2}(1-2^{-2\epsilon+1})\zeta_R({-2\epsilon+1})
\right].
\label{F_4}
\end{equation}
also in agreement with Ref. \cite{ADantas:2023wxd}.

I evaluate next ${\tilde \zeta}_T(s)$, the Lorentz violating thermal correction to the zeta function. I change the integration variable from $t$ to $w=2{\sqrt{k^2_z+M^2}\over m\beta}t$ in Eq. (\ref{zeta_13}) , and find
\begin{equation}
{\tilde \zeta}_T(s)=-{\mu^{2s}\over \Gamma(s)}{L^2\beta\over 4\pi^{3/2}}l^{2(\epsilon - 1)}(-1)^\epsilon {(2\epsilon - 1)!!\over 2^\epsilon}\sum_{m,n}
\left({\beta m\over 2 \sqrt{k^2_z +M^2}}\right)^{s-\epsilon -{1\over 2}}
\int_0^\infty dw\, w^{s-\epsilon -3/2}e^{-\sqrt{k^2_z +M^2}\beta m(w+w^{-1})/2}.
\label{zeta_18}
\end{equation}
When considering Dirichlet boundary conditions, only the term with $n=m=1$ contributes significantly to the sum. I then do the $w$-integration using the saddle point method and find
\begin{equation}
{\tilde \zeta}_T(s)=-{\mu^{2s}\over \Gamma(s)}{L^2\beta\over 4\pi}l^{2(\epsilon - 1)}(-1)^\epsilon {(2\epsilon - 1)!!\over 2^\epsilon}
\left({\beta \over 2 \sqrt{{\pi^2\over a^2} +M^2}}\right)^{s-\epsilon -{1\over 2}}
\left({2 \over \beta \sqrt{{\pi^2\over a^2} +M^2}}\right)^{{1\over 2}}e^{-\beta\sqrt{{\pi^2\over a^2} +M^2}}.
\label{zeta_19}
\end{equation}
At this point, I use the relationship between free energy and derivative of the zeta function and take advantage of Eq. (\ref{gamma_2}), to find
\begin{equation}
{\tilde F}_T={L^2\over 2\pi a^3}\left({l\over a}\right)^{2(\epsilon - 1)}(-1)^\epsilon {(2\epsilon - 1)!!}(aT)^{\epsilon +1}
\left(\pi^2+a^2M^2\right)^{\epsilon/2}e^{-{\sqrt{\pi^2+a^2M^2}\over aT}},
\label{F_5}
\end{equation}
which is the Lorentz violating thermal correction to the Casimir energy, in the small plate distance approximation for Dirichlet boundary conditions. Notice that the dominant term is the exponential suppression term.

In the case of mixed boundary conditions, the $n$-sum of Eq. (\ref{zeta_18}) starts at $n=0$, thus the only term contributing significantly has $m=1$ and $n=0$. Therefore, I can obtain ${\tilde \zeta}_T$ for mixed boundary conditions from Eq. (\ref{zeta_19}) by replacing $\pi\over a$ with $\pi\over 2a$. With that zeta function, I find the following Lorentz violating thermal correction to the Casimir energy in the small plate distance limit
\begin{equation}
{\tilde F}_T={L^2\over 2\pi a^3}\left({l\over a}\right)^{2(\epsilon - 1)}(-1)^\epsilon {(2\epsilon - 1)!!}(aT)^{\epsilon +1}
\left({\pi^2\over 4}+a^2M^2\right)^{\epsilon/2}e^{-{\sqrt{\pi^2+4a^2M^2}\over 2aT}},
\label{F_6}
\end{equation}
notice that ${\tilde F}_T$ for mixed boundary conditions as shown above has a weaker exponential suppression than ${\tilde F}_T$ for Dirichlet boundary conditions, shown in Eq. (\ref{F_5}).

Next, I examine the high temperature limit, $T\gg M, a^{-1}$. In this case, I do a Poisson resummation of the $n$-sum and find
\begin{equation}
\sum_{n=1}^\infty e^{-k^2_z  t}=-{1\over 2}+{a\over \sqrt{\pi t}}\left({1\over 2}+\sum_{n=1}^\infty e^{-a^2n^2/t}\right),
\label{Poisson_2}
\end{equation}
for Dirichlet boundary conditions, where the $-{1\over 2}$ term can be neglected since it is independent of $a$ and, when considered in the free energy calculation, it produces the energy of a single plate \cite{Cruz:2018bqt} which is not relevant to this work. The following Poisson resummation applies to the case of mixed boundary conditions
\begin{equation}
\sum_{n=0}^\infty e^{-k^2_z  t}={a\over \sqrt{\pi t}}\left({1\over 2}+\sum_{n=1}^\infty (-1)^ne^{-a^2n^2/t}\right).
\label{Poisson_3}
\end{equation}
Notice that the difference between the Poisson resummation for Dirichlet boundary conditions and that for mixed boundary conditions is only the $(-1)^n$ factor in the sum. I continue by examining Dirichlet boundary conditions. I use the Poisson resummation shown above in Eq. (\ref{zeta_7}), and find
\begin{equation}
{\tilde \zeta}(s)={\tilde \zeta}_{T,0}(s)+{\tilde \zeta}_{T,a}(s),
\label{zeta_20}
\end{equation}
where ${\tilde \zeta}_{T,0}(s)$ is the leading term of the Lorentz violating correction to the thermal zeta function, with just a linear dependence on $a$
\begin{equation}
{\tilde \zeta}_{T,0}(s)=-{\mu^{2s}\over \Gamma(s)}{L^2a\over 8\pi^{3/2}}l^{2(\epsilon - 1)}(-1)^\epsilon {(2\epsilon - 1)!!\over 2^\epsilon}\sum_{m}
\int_0^\infty dt\, t^{s-\epsilon -3/2}e^{-(k^2_\tau +M^2)t},
\label{zeta_21}
\end{equation}
and ${\tilde \zeta}_{T,a}(s)$ is the sub-leading term of the Lorentz violating correction to the thermal zeta function, with a more complicated dependence on $a$
\begin{equation}
{\tilde \zeta}_{T,a}(s)=-{\mu^{2s}\over \Gamma(s)}{L^2a\over 4\pi^{3/2}}l^{2(\epsilon - 1)}(-1)^\epsilon {(2\epsilon - 1)!!\over 2^\epsilon}\sum_{m}
\int_0^\infty dt\, t^{s-\epsilon -3/2}e^{-(k^2_\tau +M^2)t}\sum_{n=1}^\infty e^{-a^2n^2/t}.
\label{zeta_22}
\end{equation}
Since 
\begin{equation}
\sum_{m=-\infty}^{+\infty}e^{-k^2_\tau t}=1+2\sum_{m=1}^{\infty}e^{-{4\pi^2m^2\over \beta^2} t},
\label{sum_1}
\end{equation}
I proceed as in Eqs. (\ref{mass_1}) - (\ref{zeta_17}), to find
\begin{eqnarray}
{\tilde \zeta}_{T,0}(s)&=&-{\mu^{2s}\over \Gamma(s)}{L^2a\over 4\pi^{3/2}}l^{2(\epsilon - 1)}(-1)^\epsilon {(2\epsilon - 1)!!\over 2^\epsilon}\left\{
{M^{2\epsilon+1-2s}\over 2}\Gamma(s-\epsilon - {1\over 2})+\left({2\pi\over \beta }\right)^{2\epsilon+1-2s}
\left[\Gamma(s-\epsilon - {1\over 2})\right.\right.\nonumber \\
&\times&\left.\left.\zeta_R({2s-2\epsilon-1})-{M^2\beta^2\over4\pi^2}\Gamma(s-\epsilon + {1\over 2})\zeta_R({2s-2\epsilon+1})
\right]\right\}.
\label{zeta_23}
\end{eqnarray}
Proceeding as in Eqs. 
 (\ref{gamma_2}) - (\ref{F_3}), I find ${\tilde F}_{T,0}=-\beta^{-1}{\tilde \zeta}'_{T,0}(0)$
\begin{equation}
{\tilde F}_{T,0}=-{L^2aT^{4}}{(lT)}^{2(\epsilon - 1)}(2\pi )^{2\epsilon}
\left[{\zeta_R({-2\epsilon-1})\over 2\epsilon +1}+{M^2\over8\pi^2T^2}\zeta_R({-2\epsilon+1})+{1\over 4\epsilon +2}\left({M\over 2\pi T}\right)^{2\epsilon +1}
\right],
\label{F_7}
\end{equation}
where, since $M\ll 2\pi T$, the last term is negligible for all values of $\epsilon>1$.

When examining ${\tilde \zeta}_{T,a}(s)$, I find it contains a part independent of $\beta$, ${\tilde \zeta}_{T,a,1}(s)$, and a part dependent on $\beta$, ${\tilde \zeta}_{T,a,2}(s)$. I will first evaluate 
${\tilde \zeta}_{T,a,1}(s)$ in the limit of $M\ll a^{-1}$, and then in the limit of $M\gg a^{-1}$. In the limit of $M\ll a^{-1}$, I use Eq. (\ref{mass_1}), do the integration, and obtain
\begin{eqnarray}
{\tilde \zeta}_{T,a,1}(s)&=&-{\mu^{2s}\over \Gamma(s)}{L^2a^{2(s-\epsilon)}\over 4\pi^{3/2}}l^{2(\epsilon - 1)}(-1)^\epsilon {(2\epsilon - 1)!!\over 2^\epsilon}
\nonumber \\
&\times&\left[\zeta_R(2\epsilon+1-2s)\Gamma(\epsilon+{1\over 2} -s)-M^2a^2\zeta_R(2\epsilon-1-2s)\Gamma(\epsilon-{1\over 2} -s)
\right].
\label{zeta_24}
\end{eqnarray}
When considering the $M\gg a^{-1}$ limit, I change variable of integration, neglect all terms with $n>1$, and do the integral using the saddle point method to obtain
\begin{equation}
{\tilde \zeta}_{T,a,1}(s)=-{\mu^{2s}\over \Gamma(s)}{L^2\over 4\pi}l^{2(\epsilon - 1)}(-1)^\epsilon {(2\epsilon - 1)!!\over 2^\epsilon}
\left({M\over a}\right)^{\epsilon-s}e^{-2Ma}.
\label{zeta_25}
\end{equation}
Next, I evaluate ${\tilde \zeta}_{T,a,2}(s)$ by first changing integration variable from $t$ to $w={\sqrt{k^2_\tau+M^2}\over na}t$, then neglecting all terms with $n>1$ and $m>1$ in the double sum, and then using the saddle point method to integrate over $w$, to find
\begin{equation}
{\tilde \zeta}_{T,a,2}(s)=-{\mu^{2s}\over \Gamma(s)}{L^2\over 2\pi}l^{2(\epsilon - 1)}(-1)^\epsilon {(2\epsilon - 1)!!\over 2^\epsilon}
\left({\sqrt{4\pi^2\beta^{-2}+M^2}\over a}\right)^{\epsilon-s}e^{-2a\sqrt{4\pi^2\beta^{-2}+M^2}}.
\label{zeta_26}
\end{equation}
I use now Eqs. (\ref{zeta_24}) - (\ref{zeta_26}) to obtain ${\tilde F}_{T,a}=-\beta^{-1}[{\tilde \zeta}'_{T,a,1}(0)+{\tilde \zeta}'_{T,a,2}(0)]$ in the high temperature limit and, using the following
\begin{equation}
\Gamma(n+{1\over 2})=\sqrt{\pi}{(2n-1)!!\over 2^n},
\label{gamma_4}
\end{equation}
I find that, for $M\ll a^{-1}$
\begin{eqnarray}
{\tilde F}_{T,a}&=&{L^2T\over 2\pi}l^{2(\epsilon - 1)}(-1)^\epsilon {(2\epsilon - 1)!!\over (2a)^\epsilon}\left[
\left({4\pi^2T^{2}+M^2}\right)^{\epsilon/2}e^{-2a\sqrt{4\pi^2T^{2}+M^2}}+{(2\epsilon -1)!!\over 2^{\epsilon +1}}{\zeta_R(2\epsilon +1)\over a^{\epsilon}}\right.
\nonumber \\
&-&\left.M^2{(2\epsilon -3)!!\over 2^{\epsilon}}{\zeta_R(2\epsilon -1)\over a^{\epsilon-2}}\right],
\label{F_8}
\end{eqnarray}
while, for $M\gg a^{-1}$
\begin{equation}
{\tilde F}_{T,a}={L^2T\over 2\pi}l^{2(\epsilon - 1)}(-1)^\epsilon {(2\epsilon - 1)!!\over (2a)^\epsilon}\left[
\left({4\pi^2T^{2}+M^2}\right)^{\epsilon/2}e^{-2a\sqrt{4\pi^2T^{2}+M^2}}+{M^{\epsilon}\over 2}e^{-2Ma}
\right].
\label{F_9}
\end{equation}
Notice that the last term of Eq. (\ref{F_8}) is negligible when compared to the second term, since $Ma\ll1$.

Now I examine the high temperature limit for the case of mixed boundary condition. There is no difference between ${\tilde \zeta}_{T,0}(s)$ for Dirichlet boundary conditions and ${\tilde \zeta}_{T,0}(s)$ for mixed boundary conditions, therefore ${\tilde F}_{T,0}$ for mixed boundary conditions and high temperature is given by Eq. (\ref{F_7}). The only difference between ${\tilde \zeta}_{T,a,2}(s)$ for mixed boundary conditions and ${\tilde \zeta}_{T,a,2}(s)$ for Dirichlet boundary conditions is an extra factor of $-1$. The same is true for ${\tilde \zeta}_{T,a,1}(s)$ in the limit $M\gg a^{-1}$. When evaluating ${\tilde \zeta}_{T,a,1}(s)$ in the limit $M\ll a^{-1}$, I find
\begin{equation}
{\tilde \zeta}_{T,a,1}(s)=-{\mu^{2s}\over \Gamma(s)}{L^2a^{2(s-\epsilon)}\over 4\pi^{3/2}}l^{2(\epsilon - 1)}(-1)^\epsilon {(2\epsilon - 1)!!\over 2^\epsilon}
\Gamma(\epsilon+{1\over 2} -s)\left[\sum_{n=1}^\infty(-1)^n n^{2s-2\epsilon-1}\right],
\label{zeta_27}
\end{equation}
and use the following
\begin{equation}
\sum_{n=1}^\infty(-1)^n n^{-z}=-(1-2^{1-z})\zeta_R(z),
\label{zetaR_2}
\end{equation}
to obtain
\begin{equation}
{\tilde \zeta}_{T,a,1}(s)={\mu^{2s}\over \Gamma(s)}{L^2a^{2(s-\epsilon)}\over 4\pi^{3/2}}l^{2(\epsilon - 1)}(-1)^\epsilon {(2\epsilon - 1)!!\over 2^\epsilon}
\left(1-2^{2(s-\epsilon)}\right)\zeta_R(2\epsilon+1-2s)\Gamma(\epsilon+{1\over 2} -s),
\label{zeta_28}
\end{equation}
where I do not include the term proportional to $M^2$ because, as I have shown above, it is negligible. Summarizing the high temperature limit for mixed boundary conditions, I find that ${\tilde F}_{T,0}$ is given by Eq. (\ref{F_7}), while ${\tilde F}_{T,a}$
for $M\gg a^{-1}$ is given by Eq. (\ref{F_9}) multiplied by $-1$. For $M\ll a^{-1}$, I find
\begin{equation}
{\tilde F}_{T,a}=-{L^2T\over 2\pi}l^{2(\epsilon - 1)}(-1)^\epsilon {(2\epsilon - 1)!!\over (2a)^\epsilon}\left[
\left({4\pi^2T^{2}+M^2}\right)^{\epsilon/2}e^{-2a\sqrt{4\pi^2T^{2}+M^2}}+{(2\epsilon -1)!!\over 2^{\epsilon +1}}\left(1-2^{-2\epsilon}\right){\zeta_R(2\epsilon +1)\over a^{\epsilon}}\right].
\label{F_10}
\end{equation}

Last, I investigate the large mass limit, $M\gg T, a^{-1}$. In this case, I use the Poisson resummation of Eq. (\ref{Poisson_1}) for the $m$-sum,  
the Poisson resummation of Eq. (\ref{Poisson_2}) for the $n$-sum when investigating Dirichlet boundary conditions, and that of Eq. (\ref{Poisson_3}) when investigating mixed boundary conditions. I obtain
\begin{equation}
{\tilde \zeta}(s)=-{\mu^{2s}\over \Gamma(s)}{L^2a\beta\over 4\pi^{2}}l^{2(\epsilon - 1)}(-1)^\epsilon {(2\epsilon - 1)!!\over 2^\epsilon}
\int_0^\infty dt\, t^{s-\epsilon -2}e^{-M^2t}\left({1\over 2}+\sum_{m=1}^\infty e^{-\beta^2m^2/(4t)}\right)\left({1\over 2}+\sum_{n=1}^\infty e^{-a^2n^2/t}\right),
\label{zeta_29}
\end{equation}
for Dirichlet boundary conditions, and 
\begin{equation}
{\tilde \zeta}(s)=-{\mu^{2s}\over \Gamma(s)}{L^2a\beta\over 4\pi^{2}}l^{2(\epsilon - 1)}(-1)^\epsilon {(2\epsilon - 1)!!\over 2^\epsilon}
\int_0^\infty dt\, t^{s-\epsilon -2}e^{-M^2t}\left({1\over 2}+\sum_{m=1}^\infty e^{-\beta^2m^2/(4t)}\right)\left({1\over 2}+\sum_{n=1}^\infty (-1)^ne^{-a^2n^2/t}\right),
\label{zeta_30}
\end{equation}
for mixed boundary conditions. I evaluate the integrals in the large mass limit, and find
\begin{equation}
\int_0^\infty dt\, t^{s-\epsilon -2}e^{-M^2t}=M^{(2\epsilon+2-2s)}\Gamma(s-\epsilon -1),
\label{int_1}
\end{equation}
\begin{equation}
\int_0^\infty dt\, t^{s-\epsilon -2}e^{-M^2t}\sum_{m=1}^\infty e^{-\beta^2m^2/(4t)}\simeq {\sqrt{\pi}\over M}\left({2M\over \beta}\right)^{(\epsilon -s +3/2)}e^{-M\beta},
\label{int_2}
\end{equation}
\begin{equation}
\int_0^\infty dt\, t^{s-\epsilon -2}e^{-M^2t}\sum_{n=1}^\infty e^{-a^2n^2/t}\simeq {\sqrt{\pi}\over M}\left({M\over a}\right)^{(\epsilon -s +3/2)}e^{-2Ma},
\label{int_3}
\end{equation}
\begin{equation}
\int_0^\infty dt\, t^{s-\epsilon -2}e^{-M^2t}\sum_{n=1}^\infty(-1)^n e^{-a^2n^2/t}\simeq -{\sqrt{\pi}\over M}\left({M\over a}\right)^{(\epsilon -s +3/2)}e^{-2Ma},
\label{int_4}
\end{equation}
\begin{equation}
\int_0^\infty dt\, t^{s-\epsilon -2}e^{-M^2t}\sum_{m=1}^\infty e^{-\beta^2m^2/(4t)}\sum_{n=1}^\infty e^{-a^2n^2/t}\simeq {\sqrt{\pi}\over M}\left({M\over\sqrt{a^2+ \beta^2/4}}\right)^{(\epsilon -s +3/2)}e^{-2M\sqrt{a^2+ \beta^2/4}},
\label{int_5}
\end{equation}
\begin{equation}
\int_0^\infty dt\, t^{s-\epsilon -2}e^{-M^2t}\sum_{m=1}^\infty e^{-\beta^2m^2/(4t)}\sum_{n=1}^\infty(-1)^n e^{-a^2n^2/t}\simeq -{\sqrt{\pi}\over M}\left({M\over\sqrt{a^2+ \beta^2/4}}\right)^{(\epsilon -s +3/2)}e^{-2M\sqrt{a^2+ \beta^2/4}}.
\label{int_6}
\end{equation}
I use these integrals to evaluate the Lorentz violating correction to the free energy, ${\tilde F}=\beta^{-1}{\tilde \zeta}'(0)$, and obtain
\begin{equation}
{\tilde F}={L^2aM^{(\epsilon +1/2)}\over 4\pi^{3/2}}l^{2(\epsilon - 1)}(-1)^\epsilon {(2\epsilon - 1)!!\over 2^\epsilon}
\left[{(2T)^{(\epsilon+3/2)}\over 2}e^{-{M\over T}}+{e^{-2Ma}\over 2a^{(\epsilon +3/2)}}+\left({2T\over \sqrt{4a^2T^2+1}}\right)^{(\epsilon+3/2)}e^{-{M\over T}\sqrt{4a^2T^2+1}}
\right],
\label{zeta_31}
\end{equation}
for the case of Dirichlet boundary conditions, and 
\begin{equation}
{\tilde F}={L^2aM^{(\epsilon +1/2)}\over 4\pi^{3/2}}l^{2(\epsilon - 1)}(-1)^\epsilon {(2\epsilon - 1)!!\over 2^\epsilon}
\left[{(2T)^{(\epsilon+3/2)}\over 2}e^{-{M\over T}}-{e^{-2Ma}\over 2a^{(\epsilon +3/2)}}-\left({2T\over \sqrt{4a^2T^2+1}}\right)^{(\epsilon+3/2)}e^{-{M\over T}\sqrt{4a^2T^2+1}}
\right],
\label{zeta_32}
\end{equation}
for the case of mixed boundary conditions. Notice that I neglected the term coming from the integral of Eq. (\ref{int_1}) since it produces a uniform energy density term, independent of the temperature, that does not contribute to the Casimir pressure. Notice also that Eqs. (\ref{zeta_31}) and (\ref{zeta_32}), in the zero temperature limit, agree with the results of Ref. \cite{ADantas:2023wxd}.

\section{Free energy for $u^\mu$ perpendicular to the plates}
\label{4}
In this section I will obtain the Lorentz violating correction to the free energy for $u^\mu$ perpendicular to the plates, in the three limits of small plate distance, high temperature and large mass.

I start from the zeta function of Eq. (\ref{zeta_10}) and use it to investigate first the small plate distance limit. I insert into it the Poisson resummation of Eq. (\ref{Poisson_1}), and obtain
\begin{equation}
{\tilde \zeta}(s)=-{\mu^{2s}\over \Gamma(s)}{L^2\beta\over 4\pi^{3/2}}l^{2(\epsilon - 1)}(-1)^\epsilon \sum_{n}k_z^{2\epsilon}
\int_0^\infty dt\, t^{s -3/2}e^{-(k^2_z+M^2) t}\left({1\over 2}+\sum_{m=1}^\infty e^{-\beta^2m^2/4t}\right).
\label{zeta_33}
\end{equation}
I evaluate the integrals in the small plate distance limit, and find
\begin{equation}
\int_0^\infty dt\, t^{s -3/2}e^{-(k^2_z+M^2) t}\simeq k_z^{1-2s}\Gamma(s-{1\over 2})-M^2k_z^{-1-2s}\Gamma(s+{1\over 2}),
\label{int_7}
\end{equation}
where I used Eq. (\ref{mass_1}) to approximate $e^{-M^2t}$ in the small plate distance limit, and
\begin{equation}
 \sum_{n}k_z^{2\epsilon}
\int_0^\infty dt\, t^{s -3/2}e^{-(k^2_z+M^2) t}\sum_{m=1}^\infty e^{-\beta^2m^2/4t}\simeq \sqrt{\pi}
\left({\pi\over a}\right)^{2\epsilon}\left({2\over \beta}\right)\left({\beta\over 2\sqrt{{\pi^2\over a^2}+M^2}}\right)^s
e^{-\beta\sqrt{{\pi^2\over a^2}+M^2}},
\label{int_8}
\end{equation}
when $k_z$ obeys Dirichlet boundary conditions, while
\begin{equation}
 \sum_{n}k_z^{2\epsilon}
\int_0^\infty dt\, t^{s -3/2}e^{-(k^2_z+M^2) t}\sum_{m=1}^\infty e^{-\beta^2m^2/4t}\simeq \sqrt{\pi}
\left({\pi\over 2a}\right)^{2\epsilon}\left({2\over \beta}\right)\left({\beta\over 2\sqrt{{\pi^2\over 4a^2}+M^2}}\right)^se^{-\beta\sqrt{{\pi^2\over 4a^2}+M^2}},
\label{int_9}
\end{equation}
when $k_z$ obeys mixed boundary conditions. These last two integrals have been obtained by changing variable of integration and then integrating using the saddle point method, as described in the previous section. Using these integration results, the zeta function becomes
\begin{eqnarray}
{\tilde \zeta}(s)&=&-{\mu^{2s}\over \Gamma(s)}{L^2\beta\over 4\pi^{3/2}}l^{2(\epsilon - 1)}(-1)^\epsilon \left[{1\over 2}\sum_n k_z^{2\epsilon+1-2s}\Gamma(s-{1\over 2})-{M^2\over 2}\sum_nk_z^{2\epsilon-1-2s}\Gamma(s+{1\over 2})\right.
\nonumber \\
&+&\left.\sqrt{\pi}
\left({\pi\over a}\right)^{2\epsilon}\left({2\over \beta}\right)\left({\beta\over 2\sqrt{{\pi^2\over a^2}+M^2}}\right)^se^{-\beta\sqrt{{\pi^2\over a^2}+M^2}}
\right],
\label{zeta_34}
\end{eqnarray}
for Dirichlet boundary conditions, and 
\begin{eqnarray}
{\tilde \zeta}(s)&=&-{\mu^{2s}\over \Gamma(s)}{L^2\beta\over 4\pi^{3/2}}l^{2(\epsilon - 1)}(-1)^\epsilon \left[{1\over 2}\sum_n k_z^{2\epsilon+1-2s}\Gamma(s-{1\over 2})-{M^2\over 2}\sum_nk_z^{2\epsilon-1-2s}\Gamma(s+{1\over 2})\right.
\nonumber \\
&+&\left.\sqrt{\pi}
\left({\pi\over 2a}\right)^{2\epsilon}\left({2\over \beta}\right)\left({\beta\over 2\sqrt{{\pi^2\over 4a^2}+M^2}}\right)^se^{-\beta\sqrt{{\pi^2\over 4a^2}+M^2}}
\right],
\label{zeta_35}
\end{eqnarray}
for mixed boundary conditions. I evaluate the two infinite sums in the equations above using Eqs. (\ref{zeta_15}) and (\ref{zeta_16}) and the explicit form of the gamma function, to obtain the Lorentz violating correction to the free energy, ${\tilde F}=\beta^{-1}{\tilde \zeta}'(0)$, as
\begin{equation}
{\tilde F}=-{L^2\over 4a}l^{2(\epsilon - 1)}(-1)^\epsilon \left({\pi\over a}\right)^{2\epsilon}\left[\zeta_R(-2\epsilon-1)+{M^2a^2\over 2\pi^2}\zeta_R(-2\epsilon+1)-
{2aT\over \pi}e^{-{\sqrt{{\pi^2}+M^2a^2}\over aT}}
\right],
\label{F_11}
\end{equation}
for Dirichlet boundary conditions, and 
\begin{equation}
{\tilde F}={L^2\over 4a}l^{2(\epsilon - 1)}(-1)^\epsilon \left({\pi\over a}\right)^{2\epsilon}\left[(1-2^{-2\epsilon-1})\zeta_R(-2\epsilon-1)+{M^2a^2\over 2\pi^2}(1-2^{-2\epsilon+1})\zeta_R(-2\epsilon+1)+
{2aT\over 2^{2\epsilon}\pi}e^{-{\sqrt{{\pi^2}+4M^2a^2}\over 2aT}}
\right],
\label{F_12}
\end{equation}
for mixed boundary conditions. In the last equation, I used Eq. (\ref{zh_1}) to write the Hurwitz zeta function in terms of the Riemann zeta function. The zero temperature limit of the two equations above agrees with the results of Ref. \cite{ADantas:2023wxd}.

Next, I investigate the high temperature limit. In this limit a Poisson resummation of the $n$-series is needed and must be handled with care, as I show in Ref. \cite{Erdas:2025gbv}. Following that recent paper, I find
\begin{equation}
\sum_{n=1}^\infty (-1)^\epsilon k_z^{2\epsilon}e^{-k^2_z  t}=\left.t^{-\epsilon}\left({\partial\over\partial b}\right)^\epsilon{a\over \sqrt{\pi bt}}\left({1\over 2}+\sum_{n=1}^\infty e^{-{a^2n^2\over bt}}\right)\right\vert_{b=1},
\label{Poisson_4}
\end{equation}
for Dirichlet boundary conditions, where, as I did in the previous section, I neglected the term $-{1\over 2}$ since it is independent of $a$ and it produces the energy of a single plate which is not relevant to this work. I obtain the following Poisson resummation for the case of mixed boundary conditions
\begin{equation}
\sum_{n=0}^\infty (-1)^\epsilon k_z^{2\epsilon}e^{-k^2_z  t}=\left.t^{-\epsilon}\left({\partial\over\partial b}\right)^\epsilon{a\over \sqrt{\pi bt}}\left({1\over 2}+\sum_{n=1}^\infty (-1)^ne^{-{a^2n^2\over bt}}\right)\right\vert_{b=1}.
\label{Poisson_5}
\end{equation}
I use the following
\begin{equation}
\left({\partial\over\partial b}\right)^\epsilon{1\over \sqrt{ b}}=(-1)^\epsilon{(2\epsilon -1)!!\over 2^\epsilon},
\label{Poisson_6}
\end{equation}
and 
\begin{equation}
\left.\left({\partial\over\partial b}\right)^\epsilon{1\over \sqrt{ b}}e^{-{a^2n^2\over bt}}\right\vert_{b=1}\simeq \left({n^2a^2\over t}\right)^\epsilon e^{-{a^2n^2\over t}},
\label{Poisson_7}
\end{equation}
valid for $n^2a^2/t\gg 1$, to find the following Poisson resummations
\begin{equation}
\sum_{n=1}^\infty (-1)^\epsilon k_z^{2\epsilon}e^{-k^2_z  t}=t^{-\epsilon-1/2}{a\over \sqrt{\pi}}\left[(-1)^\epsilon{(2\epsilon -1)!!\over 2^{\epsilon+1}}+\sum_{n=1}^\infty  \left({n^2a^2\over t}\right)^\epsilon e^{-{a^2n^2\over t}}\right],
\label{Poisson_8}
\end{equation}
valid for Dirichlet boundary conditions, and 
\begin{equation}
\sum_{n=0}^\infty (-1)^\epsilon k_z^{2\epsilon}e^{-k^2_z  t}=t^{-\epsilon-1/2}{a\over \sqrt{\pi}}\left[(-1)^\epsilon{(2\epsilon -1)!!\over 2^{\epsilon+1}}+\sum_{n=1}^\infty  (-1)^n\left({n^2a^2\over t}\right)^\epsilon e^{-{a^2n^2\over t}}\right],
\label{Poisson_9}
\end{equation}
valid for mixed boundary conditions. I insert the resummation of Eq. (\ref{Poisson_8}) inside Eq. (\ref{zeta_10}) and, for Dirichlet boundary conditions, find
\begin{equation}
{\tilde \zeta}(s)=-{\mu^{2s}\over \Gamma(s)}{L^2a\over 4\pi^{3/2}}l^{2(\epsilon - 1)}
\int_0^\infty dt\, t^{s -\epsilon-3/2}e^{-M^2 t}\left[(-1)^\epsilon{(2\epsilon -1)!!\over 2^{\epsilon+1}}+\sum_{n=1}^\infty  \left({n^2a^2\over t}\right)^\epsilon e^{-{a^2n^2\over t}}\right]\left(1+2\sum_{m=1}^{\infty}e^{-{4\pi^2m^2\over \beta^2} t}\right).
\label{zeta_36}
\end{equation}
The four relevant integrals are evaluated below
\begin{equation}
\int_0^\infty dt\, t^{s-\epsilon -3/2}e^{-M^2t}=M^{(2\epsilon+1-2s)}\Gamma(s-\epsilon -{1\over 2}),
\label{int_10}
\end{equation}
\begin{eqnarray}
\int_0^\infty dt\, t^{s-\epsilon -3/2}e^{-M^2t}\sum_{m=1}^\infty e^{-{4\pi^2m^2\over \beta^2} t}&\simeq& \left({\beta\over 2 \pi}\right)^{2s-2\epsilon-1}
\nonumber \\
&\times&\left[\zeta_R(2s-2\epsilon -1)\Gamma(s-\epsilon-{1\over 2})-{M^2\beta^2\over 4\pi^2}\zeta_R(2s-2\epsilon +1)\Gamma(s-\epsilon+{1\over 2})
\right],
\label{int_11}
\end{eqnarray}
\begin{equation}
\int_0^\infty dt\, t^{s-2\epsilon -3/2}e^{-M^2t}\sum_{n=1}^\infty {n^{2\epsilon}} e^{-{a^2n^2\over t}}\simeq a^{2s-4\epsilon -1} \zeta_R(2\epsilon +1 -2s)\Gamma(2\epsilon +{1\over 2} -s)
\label{int_12}
\end{equation}
for $M\ll a^{-1}$, and 
\begin{equation}
\int_0^\infty dt\, t^{s-2\epsilon -3/2}e^{-M^2t}\sum_{n=1}^\infty {n^{2\epsilon}} e^{-{a^2n^2\over t}}\simeq \sqrt{\pi\over Ma}\left({a\over M}\right)^{s-2\epsilon - 1/2}e^{-2Ma}
\label{int_13}
\end{equation}
for $M\gg a^{-1}$, and finally
\begin{equation}
\int_0^\infty dt\, t^{s -2\epsilon-3/2}e^{-M^2 t}\sum_{n=1}^\infty  n^{2\epsilon} e^{-{a^2n^2\over t}}\sum_{m=1}^{\infty}e^{-{4\pi^2m^2\over \beta^2} t}\simeq {\sqrt{\pi}\over \left(a\sqrt{{4\pi^2\over\beta^2}+M^2}\right)^{1/2}}\left({a\over \sqrt{{4\pi^2\over\beta^2}+M^2}}\right)^{s-2\epsilon - 1/2}e^{-2a\sqrt{{4\pi^2\over\beta^2}+M^2}}.
\label{int_14}
\end{equation}
I use these integrals to obtain ${\tilde F}=\beta^{-1}{\tilde \zeta}'(0)$ for the case of Dirichlet boundary conditions, and find
\begin{eqnarray}
{\tilde F}&=&-{L^2aT\over 4\pi}l^{2(\epsilon - 1)}
\left[
{2\over 2\epsilon +1}\left(2\pi T\right)^{2\epsilon+1}\zeta_R(-2\epsilon -1)+\left(2\pi T\right)^{2\epsilon+1}{M^2\over 4\pi^2T^2}\zeta_R(-2\epsilon +1)+{M^{2\epsilon+1}\over 2\epsilon+1}\right.
\nonumber \\
&-&\left.a^{-2\epsilon -1} \zeta_R(2\epsilon +1){(4\epsilon -1)!!\over 2^{2\epsilon}}
-{2\over a}(4\pi^2T^2+M^2)^\epsilon e^{-2a\sqrt{4\pi^2T^2+M^2}}\right],
\label{F_13}
\end{eqnarray}
for $M\ll a^{-1}$, and 
\begin{eqnarray}
{\tilde F}&=&-{L^2aT\over 4\pi}l^{2(\epsilon - 1)}
\left[
{2\over 2\epsilon +1}\left(2\pi T\right)^{2\epsilon+1}\zeta_R(-2\epsilon -1)+\left(2\pi T\right)^{2\epsilon+1}{M^2\over 4\pi^2T^2}\zeta_R(-2\epsilon +1)+{M^{2\epsilon+1}\over 2\epsilon+1}\right.
\nonumber \\
&-&\left. {M^{2\epsilon}\over a}e^{-2Ma}
-{2\over a}(4\pi^2T^2+M^2)^\epsilon e^{-2a\sqrt{4\pi^2T^2+M^2}}\right],
\label{F_14}
\end{eqnarray}
valid for $M\gg a^{-1}$. Notice how, in Eq. (\ref{F_13}), the term proportional to $M^{2\epsilon +1}$ is negligible, since $M\ll a^{-1}\ll T$, and, in Eq. (\ref{F_14}), the term proportional to $e^{-2Ma}$ is negligible, since $a^{-1}\ll M\ll T$. It is interesting to notice that the first three terms of the last two equations, which include the leading term of the Lorentz violating contribution to the free energy in the high temperature limit, are identical to ${\tilde F}_{T,0}$ of Eq. (\ref{F_7}) from the previous section, which  indicates that, to leading order, the high temperature limit of ${\tilde F}$, is the same when $u^\mu$ is parallel to the plates and when $u^\mu$ is perpendicular to the plates.

To examine the case of mixed boundary conditions, I insert the resummation of Eq. (\ref{Poisson_9}) into the zeta function of Eq. (\ref{zeta_10}) and find
\begin{eqnarray}
{\tilde \zeta}(s)&=&-{\mu^{2s}\over \Gamma(s)}{L^2a\over 4\pi^{3/2}}l^{2(\epsilon - 1)}
\int_0^\infty dt\, t^{s -\epsilon-3/2}e^{-M^2 t}
\nonumber \\
&\times&\left[(-1)^\epsilon{(2\epsilon -1)!!\over 2^{\epsilon+1}}+\sum_{n=1}^\infty  (-1)^n\left({n^2a^2\over t}\right)^\epsilon e^{-{a^2n^2\over t}}\right]\left(1+2\sum_{m=1}^{\infty}e^{-{4\pi^2m^2\over \beta^2} t}\right).
\label{zeta_37}
\end{eqnarray}
I need three more integrals to continue, and list them below
\begin{equation}
\int_0^\infty dt\, t^{s-2\epsilon -3/2}e^{-M^2t}\sum_{n=1}^\infty (-1)^n{n^{2\epsilon}} e^{-{a^2n^2\over t}}\simeq -(1-2^{2s-2\epsilon})a^{2s-4\epsilon -1} \zeta_R(2\epsilon +1 -2s)\Gamma(2\epsilon +{1\over 2} -s),
\label{int_15}
\end{equation}
for $M\ll a^{-1}$, and 
\begin{equation}
\int_0^\infty dt\, t^{s-2\epsilon -3/2}e^{-M^2t}\sum_{n=1}^\infty (-1)^n{n^{2\epsilon}} e^{-{a^2n^2\over t}}\simeq -\sqrt{\pi\over Ma}\left({a\over M}\right)^{s-2\epsilon - 1/2}e^{-2Ma},
\label{int_16}
\end{equation}
for $M\gg a^{-1}$, and last
\begin{eqnarray}
\int_0^\infty dt\, t^{s -2\epsilon-3/2}e^{-M^2 t}\sum_{n=1}^\infty (-1)^n n^{2\epsilon} e^{-{a^2n^2\over t}}\sum_{m=1}^{\infty}e^{-{4\pi^2m^2\over \beta^2} t}&\simeq& -{\sqrt{\pi}\over \left(a\sqrt{{4\pi^2\over\beta^2}+M^2}\right)^{1/2}}
\nonumber \\
&\times&\left({a\over \sqrt{{4\pi^2\over\beta^2}+M^2}}\right)^{s-2\epsilon - 1/2}e^{-2a\sqrt{{4\pi^2\over\beta^2}+M^2}}.
\label{int_17}
\end{eqnarray}
I use these integrals, and the ones listed in the previous paragraph, to obtain the high temperature limit of the Lorentz violating correction to the free energy, ${\tilde F}$, for mixed boundary conditions
\begin{eqnarray}
{\tilde F}&=&-{L^2aT\over 4\pi}l^{2(\epsilon - 1)}
\left[
{2\over 2\epsilon +1}\left(2\pi T\right)^{2\epsilon+1}\zeta_R(-2\epsilon -1)+\left(2\pi T\right)^{2\epsilon+1}{M^2\over 4\pi^2T^2}\zeta_R(-2\epsilon +1)+{M^{2\epsilon+1}\over 2\epsilon+1}\right.
\nonumber \\
&+&\left.a^{-2\epsilon -1} (1-2^{-2\epsilon})\zeta_R(2\epsilon +1){(4\epsilon -1)!!\over 2^{2\epsilon}}
+{2\over a}(4\pi^2T^2+M^2)^\epsilon e^{-2a\sqrt{4\pi^2T^2+M^2}}\right],
\label{F_15}
\end{eqnarray}
for $M\ll a^{-1}$, and 
\begin{eqnarray}
{\tilde F}&=&-{L^2aT\over 4\pi}l^{2(\epsilon - 1)}
\left[
{2\over 2\epsilon +1}\left(2\pi T\right)^{2\epsilon+1}\zeta_R(-2\epsilon -1)+\left(2\pi T\right)^{2\epsilon+1}{M^2\over 4\pi^2T^2}\zeta_R(-2\epsilon +1)+{M^{2\epsilon+1}\over 2\epsilon+1}\right.
\nonumber \\
&+&\left. {M^{2\epsilon}\over a}e^{-2Ma}
+{2\over a}(4\pi^2T^2+M^2)^\epsilon e^{-2a\sqrt{4\pi^2T^2+M^2}}\right],
\label{F_16}
\end{eqnarray}
valid for $M\gg a^{-1}$. Notice that, in Eq. (\ref{F_15}), the term proportional to $M^{2\epsilon +1}$ is negligible and, in Eq. (\ref{F_16}), the term proportional to $e^{-2Ma}$ is negligible, as I explained in the paragraph above.

Last, I investigate the large mass limit, $M\gg T, a^{-1}$. In this case, I do a Poisson resummation of the $m$ and $n$ series and, for the case of Dirichlet boundary conditions, find 
\begin{equation}
{\tilde \zeta}(s)=-{\mu^{2s}\over \Gamma(s)}{L^2a\beta\over 4\pi^{2}}l^{2(\epsilon - 1)}
\int_0^\infty dt\, t^{s -\epsilon-2}e^{-M^2 t}\left[(-1)^\epsilon{(2\epsilon -1)!!\over 2^{\epsilon+1}}+\sum_{n=1}^\infty  \left({n^2a^2\over t}\right)^\epsilon e^{-{a^2n^2\over t}}\right]\left({1\over 2}+\sum_{m=1}^{\infty}e^{-{\beta^2m^2\over 4t } }\right),
\label{zeta_38}
\end{equation}
while, for mixed boundary conditions, I obtain
\begin{equation}
{\tilde \zeta}(s)=-{\mu^{2s}\over \Gamma(s)}{L^2a\beta\over 4\pi^{2}}l^{2(\epsilon - 1)}
\int_0^\infty dt\, t^{s -\epsilon-2}e^{-M^2 t}\left[(-1)^\epsilon{(2\epsilon -1)!!\over 2^{\epsilon+1}}+\sum_{n=1}^\infty  (-1)^n\left({n^2a^2\over t}\right)^\epsilon e^{-{a^2n^2\over t}}\right]\left({1\over 2}+\sum_{m=1}^{\infty}e^{-{\beta^2m^2\over 4t } }\right).
\label{zeta_39}
\end{equation}
Once I do the integrals using Eqs. (\ref{int_1}) - (\ref{int_6}) or similar and neglect a uniform energy density term that does not depend on temperature, I find
the Lorentz violating part of the free energy ${\tilde F}$ for Dirichlet boundary conditions in the large mass limit
\begin{equation}
{\tilde F}={L^2\over 4\pi^{3/2}}l^{2(\epsilon - 1)}
\left[M^{2\epsilon + {1\over2}} {e^{-2Ma}\over 2a^{1\over 2}}
+{\sqrt{2}}(-1)^\epsilon{(2\epsilon -1)!!\over 2^{\epsilon+1}}aT(MT)^{\epsilon +{1\over2}}e^{-M/T}
+\sqrt{2T\over M}{(2MTa)^{2\epsilon+1}\over (4a^2T^2+1)^{\epsilon + {3\over 4}}}{e^{-{M\over T}\sqrt{4a^2T^2+1}}}
\right],
\label{F_17}
\end{equation}
and for mixed boundary conditions
\begin{equation}
{\tilde F}=-{L^2\over 4\pi^{3/2}}l^{2(\epsilon - 1)}
\left[M^{2\epsilon + {1\over2}} {e^{-2Ma}\over 2a^{1\over 2}}
-{\sqrt{2}}(-1)^\epsilon{(2\epsilon -1)!!\over 2^{\epsilon+1}}aT(MT)^{\epsilon +{1\over2}}e^{-M/T}
+\sqrt{2T\over M}{(2MTa)^{2\epsilon+1}\over (4a^2T^2+1)^{\epsilon + {3\over 4}}}{e^{-{M\over T}\sqrt{4a^2T^2+1}}}
\right].
\label{F_18}
\end{equation}
These last two equations, in the limit of zero temperature, agree with the results of Ref. \cite{ADantas:2023wxd}.

\section{Casimir pressure}
\label{5}
The Casimir pressure is defined as
\begin{equation}
P_C=-{1\over L^2}{\partial F \over \partial a}
\label{PC_01}
\end{equation}
where $F$ is the free energy of Eq. (\ref{F_2}). The Casimir pressure without Lorentz violation, $\bar{P}_C$, is well known and can be found, for example, in Refs. \cite{Fierz:1960zq,Mehra:1967wf,CougoPinto:1998jg,Erdas:2013jga,Erdas:2013dha}. In this section I will focus on the Lorentz violating corrections to the Casimir pressure, ${\tilde P}_C$.

I begin by examining the situation where $u_\mu$ is parallel to the plates and the scalar field satisfies Dirichlet boundary conditions at the plates. The Lorentz violating correction to the Casimir pressure in the small plate distance limit is given by
\begin{eqnarray}
{\tilde P}_C&=&-{1\over 4 a^4 } 
\left(l\over a\right)^{2(\epsilon - 1)}
\left[\pi^{2\epsilon}\zeta_R(-2\epsilon -1)+(2\epsilon -1)\pi^{2(\epsilon - 1)}{M^2a^2\over 2}\zeta_R(1-2\epsilon )\right.
\nonumber \\
&+&
\left.2{(-1)^\epsilon {(2\epsilon - 1)!!}\over \pi}(aT)^{\epsilon}
\left(\pi^2+a^2M^2\right)^{(\epsilon+1)/2}e^{-{\sqrt{\pi^2+a^2M^2}\over aT}}\right],
\label{PC_02}
\end{eqnarray}
where the temperature correction weakens the plates attraction when $\epsilon$ is odd. In the high temperature regime, ${\tilde P}_C$ is given by
\begin{eqnarray}
{\tilde P}_C&=&{T^{4}}{(lT)}^{2(\epsilon - 1)}(2\pi )^{2\epsilon}
\left[{\zeta_R({-2\epsilon-1})\over 2\epsilon +1}+{M^2\over8\pi^2T^2}\zeta_R({-2\epsilon+1})
\right]
\nonumber \\
&+&{T\over \pi}l^{2(\epsilon - 1)}(-1)^\epsilon {(2\epsilon - 1)!!\over (2a)^\epsilon}\left[
\left({4\pi^2T^{2}+M^2}\right)^{(\epsilon+1)/2}e^{-2a\sqrt{4\pi^2T^{2}+M^2}}+{\epsilon(2\epsilon -1)!!}{\zeta_R(2\epsilon +1)\over (2a)^{\epsilon+1}}\right],
\label{PC_03}
\end{eqnarray}
valid for $M\ll a^{-1}$, and 
\begin{eqnarray}
{\tilde P}_C&=&{T^{4}}{(lT)}^{2(\epsilon - 1)}(2\pi )^{2\epsilon}
\left[{\zeta_R({-2\epsilon-1})\over 2\epsilon +1}+{M^2\over8\pi^2T^2}\zeta_R({-2\epsilon+1})
\right]
\nonumber \\
&+&{T\over \pi}l^{2(\epsilon - 1)}(-1)^\epsilon {(2\epsilon - 1)!!\over (2a)^\epsilon}\left[
\left({4\pi^2T^{2}+M^2}\right)^{(\epsilon+1)/2}e^{-2a\sqrt{4\pi^2T^{2}+M^2}}+{M^{\epsilon +1}\over 2}e^{-2Ma}\right],
\label{PC_04}
\end{eqnarray}
valid for $M\gg a^{-1}$. In both of these equations the dominant term is the Stefan-Boltzmann-like term, proportional to $T^{2(\epsilon+1)}$. When I examine the large mass limit, I find
\begin{eqnarray}
{\tilde P}_C&=&{M^{(\epsilon +1/2)}\over 4\pi^{3/2}}l^{2(\epsilon - 1)}(-1)^\epsilon {(2\epsilon - 1)!!\over 2^\epsilon}
\left[M{e^{-2Ma}\over a^{(\epsilon +1/2)}}-{(2T)^{(\epsilon+3/2)}\over 2}e^{-{M\over T}}\right.
\nonumber \\
&-&
\left.\left(1-{4a^2MT\over \sqrt{4a^2T^2+1}}\right)
\left({2T\over \sqrt{4a^2T^2+1}}\right)^{(\epsilon +3/2)}e^{-{M\over T}\sqrt{4a^2T^2+1}}
\right],
\label{PC_05}
\end{eqnarray}
repulsive and displaying again a dominant exponential suppression term.

In the scenario where $u_\mu$ is parallel to the plates and the scalar field satisfies mixed boundary conditions at the plates, I find that the Lorentz violating correction to the Casimir pressure in the small plate distance limit is given by
\begin{eqnarray}
{\tilde P}_C&=&{1\over 4 a^4 } 
\left(l\over a\right)^{2(\epsilon - 1)}
\left[\pi^{2\epsilon}(1-2^{-2\epsilon-1})\zeta_R(-2\epsilon -1)+(2\epsilon -1)\pi^{2(\epsilon - 1)}(1-2^{-2\epsilon+1}){M^2a^2\over 2}\zeta_R(1-2\epsilon )\right.
\nonumber \\
&-&
\left.2{(-1)^\epsilon {(2\epsilon - 1)!!}\over \pi}(aT)^{\epsilon}
\left({\pi^2\over 4}+a^2M^2\right)^{(\epsilon+1)/2}e^{-{\sqrt{\pi^2+4a^2M^2}\over 2aT}}\right],
\label{PC_06}
\end{eqnarray}
notice how, in the small plate distance limit, the Lorentz violating correction to the Casimir pressure for mixed boundary conditions has opposite sign of the Lorentz violating correction for Dirichlet boundary conditions.
The Lorentz violating correction to the Casimir pressure in the high temperature limit under mixed boundary conditions is given by
\begin{eqnarray}
{\tilde P}_C&=&{T^{4}}{(lT)}^{2(\epsilon - 1)}(2\pi )^{2\epsilon}
\left[{\zeta_R({-2\epsilon-1})\over 2\epsilon +1}+{M^2\over8\pi^2T^2}\zeta_R({-2\epsilon+1})
\right]
\nonumber \\
&-&{T\over \pi}l^{2(\epsilon - 1)}(-1)^\epsilon {(2\epsilon - 1)!!\over (2a)^\epsilon}\left[
\left({4\pi^2T^{2}+M^2}\right)^{(\epsilon+1)/2}e^{-2a\sqrt{4\pi^2T^{2}+M^2}}+{\epsilon(2\epsilon -1)!!}(1-2^{-2\epsilon}){\zeta_R(2\epsilon +1)\over (2a)^{\epsilon+1}}\right],
\label{PC_07}
\end{eqnarray}
valid for $M\ll a^{-1}$, and 
\begin{eqnarray}
{\tilde P}_C&=&{T^{4}}{(lT)}^{2(\epsilon - 1)}(2\pi )^{2\epsilon}
\left[{\zeta_R({-2\epsilon-1})\over 2\epsilon +1}+{M^2\over8\pi^2T^2}\zeta_R({-2\epsilon+1})
\right]
\nonumber \\
&-&{T\over \pi}l^{2(\epsilon - 1)}(-1)^\epsilon {(2\epsilon - 1)!!\over (2a)^\epsilon}\left[
\left({4\pi^2T^{2}+M^2}\right)^{(\epsilon+1)/2}e^{-2a\sqrt{4\pi^2T^{2}+M^2}}+{M^{\epsilon +1}\over 2}e^{-2Ma}\right],
\label{PC_08}
\end{eqnarray}
valid for $M\gg a^{-1}$. Notice that the dominant term is the Stefan-Boltzmann-like term, proportional to $T^{2(\epsilon+1)}$, as it is for Dirichlet boundary conditions and, in both cases of mixed and Dirichlet boundary conditions, the Lorentz violating correction to the Casimir pressure has the same sign but it is stronger in the case of Dirichlet boundary conditions. The large mass limit of ${\tilde P}_C$ under mixed boundary conditions is
\begin{eqnarray}
{\tilde P}_C&=&-{M^{(\epsilon +1/2)}\over 4\pi^{3/2}}l^{2(\epsilon - 1)}(-1)^\epsilon {(2\epsilon - 1)!!\over 2^\epsilon}
\left[M{e^{-2Ma}\over a^{(\epsilon +1/2)}}+{(2T)^{(\epsilon+3/2)}\over 2}e^{-{M\over T}}\right.
\nonumber \\
&-&
\left.\left(1-{4a^2MT\over \sqrt{4a^2T^2+1}}\right)
\left({2T\over \sqrt{4a^2T^2+1}}\right)^{(\epsilon +3/2)}e^{-{M\over T}\sqrt{4a^2T^2+1}}
\right],
\label{PC_09}
\end{eqnarray}
with a dominant exponential suppression term but, opposite to the case of Dirichlet boundary conditions,  attractive.

Last, I examine the situation where $u_\mu$ is perpendicular to the plates. In the case of Dirichlet boundary conditions, I find
\begin{equation}
{\tilde P}_C
=-{1\over 4a^2}l^{2(\epsilon - 1)}(-1)^\epsilon \left({\pi\over a}\right)^{2\epsilon}\left[(2\epsilon+1)\zeta_R(-2\epsilon-1)+(2\epsilon-1){M^2a^2\over 2\pi^2}\zeta_R(-2\epsilon+1)-
{2\over \pi}e^{-{\sqrt{{\pi^2}+M^2a^2}\over aT}}\sqrt{{\pi^2}+M^2a^2}
\right],
\label{PC_10}
\end{equation}
in the small plate distance limit, 
\begin{eqnarray}
{\tilde P}_C&=&{T\over 4\pi}l^{2(\epsilon - 1)}
\left[
{2\over 2\epsilon +1}\left(2\pi T\right)^{2\epsilon+1}\zeta_R(-2\epsilon -1)+\left(2\pi T\right)^{2\epsilon+1}{M^2\over 4\pi^2T^2}\zeta_R(-2\epsilon +1)\right.
\nonumber \\
&+&\left.a^{-2\epsilon -1} \zeta_R(2\epsilon +1)\epsilon{(4\epsilon -1)!!\over 2^{2\epsilon-1}}
+{4}(4\pi^2T^2+M^2)^{\epsilon+{1\over 2}} e^{-2a\sqrt{4\pi^2T^2+M^2}}\right],
\label{PC_11}
\end{eqnarray}
in the high temperature limit for $M\ll a^{-1}$, 
\begin{eqnarray}
{\tilde P}_C&=&{T\over 4\pi}l^{2(\epsilon - 1)}
\left[{2\over 2\epsilon +1}
\left(2\pi T\right)^{2\epsilon+1}\zeta_R(-2\epsilon -1)+\left(2\pi T\right)^{2\epsilon+1}{M^2\over 4\pi^2T^2}\zeta_R(-2\epsilon +1)
\right.
\nonumber \\
&+&\left. 
2{M^{2\epsilon+1}}e^{-2Ma}
+4(4\pi^2T^2+M^2)^{\epsilon+{1\over 2}} e^{-2a\sqrt{4\pi^2T^2+M^2}}
\right],
\label{PC_12}
\end{eqnarray}
in the high temperature limit for $M\gg a^{-1}$, and 
\begin{eqnarray}
{\tilde P}_C&=&{1\over 4\pi^{3/2}}l^{2(\epsilon - 1)}
\left[M^{2\epsilon + {3\over2}} {e^{-2Ma}\over a^{1\over 2}}
-{\sqrt{2}}(-1)^\epsilon{(2\epsilon -1)!!\over 2^{\epsilon+1}}T(MT)^{\epsilon +{1\over2}}e^{-M/T}
\right.
\nonumber \\
&+&\left. 
\sqrt{8T^3 M}\left({4a^2MT\over \sqrt{4a^2T^2+1} }-2\epsilon-1\right){(2MTa)^{2\epsilon}\over (4a^2T^2+1)^{\epsilon + {3\over 4}}}{e^{-{M\over T}\sqrt{4a^2T^2+1}}}
\right],
\label{PC_13}
\end{eqnarray}
in the large mass limit. Notice that, in the small plate distance limit, ${\tilde P}_C$ has opposite sign than in the high temperature limit where the dominant term is the Stefan-Boltzmann-like term. In the large mass limit ${\tilde P}_C$ is repulsive  with a dominant exponential suppression term.

Finally, in the case of mixed boundary conditions, I obtain
\begin{eqnarray}
{\tilde P}_C&=&{1\over 4a^2}l^{2(\epsilon - 1)}(-1)^\epsilon \left({\pi\over a}\right)^{2\epsilon}\left[(2\epsilon+1)(1-2^{-2\epsilon-1})\zeta_R(-2\epsilon-1)+(2\epsilon-1)(1-2^{-2\epsilon+1}){M^2a^2\over 2\pi^2}\zeta_R(-2\epsilon+1)
\right.
\nonumber \\
&+&\left. 
{1\over \pi}e^{-{\sqrt{{\pi^2}+4M^2a^2}\over 2aT}}\sqrt{{\pi^2}+4M^2a^2}
\right],
\label{PC_14}
\end{eqnarray}
in the small plate distance limit,
\begin{eqnarray}
{\tilde P}_C&=&{T\over 4\pi}l^{2(\epsilon - 1)}
\left[
{2\over 2\epsilon +1}\left(2\pi T\right)^{2\epsilon+1}\zeta_R(-2\epsilon -1)+\left(2\pi T\right)^{2\epsilon+1}{M^2\over 4\pi^2T^2}\zeta_R(-2\epsilon +1)\right.
\nonumber \\
&-&\left.a^{-2\epsilon -1}(1-2^{-2\epsilon}) \zeta_R(2\epsilon +1)\epsilon{(4\epsilon -1)!!\over 2^{2\epsilon-1}}
-{4}(4\pi^2T^2+M^2)^{\epsilon+{1\over 2}} e^{-2a\sqrt{4\pi^2T^2+M^2}}\right],
\label{PC_15}
\end{eqnarray}
in the high temperature limit for $M\ll a^{-1}$, 
\begin{eqnarray}
{\tilde P}_C&=&{T\over 4\pi}l^{2(\epsilon - 1)}
\left[{2\over 2\epsilon +1}
\left(2\pi T\right)^{2\epsilon+1}\zeta_R(-2\epsilon -1)+\left(2\pi T\right)^{2\epsilon+1}{M^2\over 4\pi^2T^2}\zeta_R(-2\epsilon +1)
\right.
\nonumber \\
&-&\left. 
2{M^{2\epsilon+1}}e^{-2Ma}
-4(4\pi^2T^2+M^2)^{\epsilon+{1\over 2}} e^{-2a\sqrt{4\pi^2T^2+M^2}}
\right],
\label{PC_16}
\end{eqnarray}
in the high temperature limit for $M\gg a^{-1}$, and 
\begin{eqnarray}
{\tilde P}_C&=&-{1\over 4\pi^{3/2}}l^{2(\epsilon - 1)}
\left[M^{2\epsilon + {3\over2}} {e^{-2Ma}\over a^{1\over 2}}
+{\sqrt{2}}(-1)^\epsilon{(2\epsilon -1)!!\over 2^{\epsilon+1}}T(MT)^{\epsilon +{1\over2}}e^{-M/T}
\right.
\nonumber \\
&+&\left. 
\sqrt{8T^3 M}\left({4a^2MT\over \sqrt{4a^2T^2+1} }-2\epsilon-1\right){(2MTa)^{2\epsilon}\over (4a^2T^2+1)^{\epsilon + {3\over 4}}}{e^{-{M\over T}\sqrt{4a^2T^2+1}}}
\right],
\label{PC_17}
\end{eqnarray}
in the large mass limit. Notice how the Stefan-Boltzmann-like term appearing in Eqs. (\ref{PC_15}) and (\ref{PC_16}) is the same as in the case of Dirichlet boundary conditions, while the Lorentz violating corrections to $P_C$ for mixed boundary conditions in the limits of small plate distance and large mass have opposite signs when compared to those for Dirichlet boundary conditions under the same limits. 
\section{Discussion and conclusions}
\label{6}
In this paper I used the zeta function technique to investigate the finite temperature Casimir effect of a charged and massive Lorentz-violating scalar field. This scalar field satisfies a modified Klein-Gordon equation that breaks Lorentz invariance in a CPT-even aether-like manner, with the breaking caused by a constant space-like unit four-vector $u^\mu$ coupled to higher order derivatives of $\phi$. I studied the cases of $\phi$ satisfying Dirichlet and mixed boundary conditions on two parallel plates. I did not investigate Neumann boundary conditions because they produce the same results as Dirichlet boundary conditions.
In Sec. \ref{3}, I obtained simple analytic expressions for the free energy in the asymptotic cases of short plate distance, high temperature and large mass when $u^\mu$ is parallel to the plates, for both Dirichlet and mixed boundary conditions. In Sec. \ref{4} I did the same for the case of  $u^\mu$ perpendicular to the plates.
In Sec. \ref{5} I obtained simple analytic expressions of the Casimir pressure in the three asymptotic cases for both types of boundary conditions and for the two scenarios where $u^\mu$ is parallel and perpendicular to the plates. 

Since the Casimir pressure is the measurable quantity, I list below my results for the Lorentz violating correction to the Casimir  pressure, under Dirichlet boundary conditions, for the three asymptotic limits, the two directions of $u^\mu$, and for critical exponent $\epsilon =2,3$.

\begin{itemize}
  \item $\epsilon =2$ and $u^\mu$ parallel
  
  \underline{ Small plate distance} 
 \begin{equation}
{\tilde P}_C={\pi^{4}\over 16 a^4 } 
\!\left(l\over a\right)^{2}\!
\left[{1\over 63}-{M^2a^2\over 20\pi^2}-{24\over \pi^5}a^2T^2(\pi^2+a^2M^2)^{3/2}e^{-{\sqrt{\pi^2+a^2M^2}\over aT}}
\right]
\label{end_01}
\end{equation}

 \underline{ High temperature} 
 \begin{equation}
{\tilde P}_C=-\pi^4T^4(lT)^2\left({4\over 315}-{M^2\over 60\pi^2 T^2}\right)+
{3\over 4\pi}T\left({l\over a}\right)^2\left[(4\pi^2T^2+M^2)^{3/2}e^{-2a\sqrt{4\pi^2T^2+M^2}}+{3\zeta_R(5)\over 4 a^3}\right]
\label{end_02}
\end{equation}
valid for $M\ll a^{-1}$ and where $\zeta_R(5)=1.03693$,

 \begin{equation} 
{\tilde P}_C=-\pi^4T^4(lT)^2\left({4\over 315}-{M^2\over 60\pi^2 T^2}\right)+
{3\over 4\pi}T\left({l\over a}\right)^2\left[(4\pi^2T^2+M^2)^{3/2}e^{-2a\sqrt{4\pi^2T^2+M^2}}+{M^3\over 2}e^{-2Ma}\right]
\label{end_02b}
\end{equation}
valid for $M\gg a^{-1}$.

 \underline{ Large mass} 
 \begin{equation}
{\tilde P}_C={3{M}^{5/2}\over 16\pi^{3/2} } l^{2}\!\!  \left[{M\over a^{5/2}}e^{-2Ma}-{(2T)^{7/2}\over 2}e^{-{M\over T}}-
\left(1-{4a^2MT\over \sqrt{4a^2T^2+1}}\right)
\left({2T\over \sqrt{4a^2T^2+1}}\right)^{7/2}e^{-{M\over T}\sqrt{4a^2T^2+1}}\right].
\label{end_03}
\end{equation}

 \item $\epsilon =2$ and $u^\mu$ perpendicular
  
 \underline{ Small plate distance} 
 \begin{equation}
{\tilde P}_C={\pi^{4}\over 16 a^4 } 
\left(l\over a\right)^{2}\left[{5\over 63}-{m^2a^2\over 20\pi^2}+{8\over \pi}\sqrt{\pi^2+a^2M^2}e^{-{\sqrt{\pi^2+a^2M^2}\over aT}}\right]
\label{end_04}
\end{equation}

 \underline{ High temperature} 
 \begin{equation}
{\tilde P}_C=-\pi^4T^4(lT)^2\left[{4\over 315}-{M^2\over 60\pi^2 T^2}-{\left(4+{M^2\over\pi^2 T^2}\right)^{5/2}}e^{-2a\sqrt{4\pi^2T^2+M^2}}-{105\over 16}{\zeta_R(5)\over (\pi a T)^5}\right]
\label{end_05}
\end{equation}
valid for $M\ll a^{-1}$,

\begin{equation} 
{\tilde P}_C=-\pi^4T^4(lT)^2\left[{4\over 315}-{M^2\over 60\pi^2 T^2}-{\left(4+{M^2\over\pi^2 T^2}\right)^{5/2}}e^{-2a\sqrt{4\pi^2T^2+M^2}}-{1\over 2}\left({M\over \pi T}\right)^5e^{-2Ma}\right]
\label{end_05b}
\end{equation}
valid for $M\gg a^{-1}$.

  \underline{ Large mass} 
  \begin{equation}
{\tilde P}_C={{M}^{5/2}\over 4\pi^{3/2} } l^{2}\!\!  \left[{M^3\over a^{1/2}}e^{-2Ma}-{3T^{7/2}\over 2^{5/2}}e^{-{M\over T}}+
\sqrt{8T^3}M^2\left({4a^2MT\over \sqrt{4a^2T^2+1}}-5\right)
{(2Ta)^4\over (4a^2T^2+1)^{11/4}}e^{-{M\over T}\sqrt{4a^2T^2+1}}\right].
\label{end_06}
\end{equation}

 \item $\epsilon =3$ and $u^\mu$ parallel
  
  \underline{ Small plate distance} 
  \begin{equation}
{\tilde P}_C=-{\pi^{6}\over 96 a^4 } 
\!\left(l\over a\right)^{4}\!
\left[{1\over 10}-{5M^2a^2\over 21\pi^2}-{720\over \pi^7}a^3T^3(\pi^2+a^2M^2)^{2}e^{-{\sqrt{\pi^2+a^2M^2}\over aT}}\right]
\label{end_07}
\end{equation}

  \underline{ High temperature} 
  \begin{equation}
{\tilde P}_C=\pi^6T^4(lT)^4\left({4\over 105}-{2M^2\over 63\pi^2 T^2}\right)-
{15\over 8\pi}Ta\left({l\over a}\right)^4\left[(4\pi^2T^2+M^2)^{2}e^{-2a\sqrt{4\pi^2T^2+M^2}}+{45\zeta_R(7)\over 16 a^4}\right]
\label{end_08}
\end{equation}
valid for $M\ll a^{-1}$ and where $\zeta_R(7)=1.00835\simeq 1$,

  \begin{equation} 
{\tilde P}_C=\pi^6T^4(lT)^4\left({4\over 105}-{2M^2\over 63\pi^2 T^2}\right)-
{15\over 8\pi}Ta\left({l\over a}\right)^4\left[(4\pi^2T^2+M^2)^{2}e^{-2a\sqrt{4\pi^2T^2+M^2}}+{M^4\over 2}e^{-2Ma}\right]
\label{end_08b}
\end{equation}
valid for $M\gg a^{-1}$.

  \underline{ Large mass} 
   \begin{equation}
{\tilde P}_C=-{15{M}^{7/2}\over 32\pi^{3/2} } l^{4}\!\!  \left[{M\over a^{7/2}}e^{-2Ma}-{(2T)^{9/2}\over 2}e^{-{M\over T}}-
\left(1-{4a^2MT\over \sqrt{4a^2T^2+1}}\right)
\left({2T\over \sqrt{4a^2T^2+1}}\right)^{9/2}e^{-{M\over T}\sqrt{4a^2T^2+1}}\right].
\label{end_09}
\end{equation}

 \item $\epsilon =3$ and $u^\mu$ perpendicular
 
  \underline{ Small plate distance} 
 \begin{equation}
{\tilde P}_C={\pi^{6}\over 96 a^4 } 
\left(l\over a\right)^{4}\left[{7\over 10}-{5M^2a^2\over 21\pi^2}-{48\over \pi}\sqrt{\pi^2+a^2M^2}e^{-{\sqrt{\pi^2+a^2M^2}\over aT}}\right]
\label{end_10}
\end{equation}

  \underline{ High temperature} 
 \begin{equation}
{\tilde P}_C=\pi^6T^4(lT)^4\left[{4\over 105}-{2M^2\over 63\pi^2 T^2}+{\left(4+{M^2\over\pi^2 T^2}\right)^{7/2}}e^{-2a\sqrt{4\pi^2T^2+M^2}}+{31,185\over 128}{\zeta_R(7)\over (\pi a T)^7}\right]
\label{end_11}
\end{equation}
valid for $M\ll a^{-1}$,

 \begin{equation} 
{\tilde P}_C=\pi^6T^4(lT)^4\left[{4\over 105}-{2M^2\over 63\pi^2 T^2}+{\left(4+{M^2\over\pi^2 T^2}\right)^{7/2}}e^{-2a\sqrt{4\pi^2T^2+M^2}}+{1\over 2}\left({M\over \pi T}\right)^7e^{-2Ma}\right]
\label{end_11b}
\end{equation}
valid for $M\gg a^{-1}$.

  \underline{ Large mass}
  \begin{equation}
{\tilde P}_C={{M}^{7/2}\over 4\pi^{3/2} } l^{4}\!\!  \left[{M^4\over a^{1/2}}e^{-2Ma}-{15T^{9/2}\over 2^{7/2}}e^{-{M\over T}}+
\sqrt{8T^3}M^3\left({4a^2MT\over \sqrt{4a^2T^2+1}}-7\right)
{(2Ta)^6\over (4a^2T^2+1)^{15/4}}e^{-{M\over T}\sqrt{4a^2T^2+1}}\right].
\label{end_12}
\end{equation}
\end{itemize}

The Casimir pressure without Lorentz violation, $\bar{P}_C$, under Dirichlet boundary conditions and for the three asymptotic limits discussed in this paper, is listed below

  \underline{ Small plate distance}
 \begin{equation}
\bar{P}_C=-{\pi^{2}\over 48 a^4 } 
\left({1\over 5}-{M^2a^2\over \pi^2}\right)+{\pi T\over a^3}e^{-{\sqrt{\pi^2+M^2a^2}\over aT}}
\label{end_13}
\end{equation}

  \underline{ High temperature} 
 \begin{equation}
\bar{P}_C={\pi^{2}T^4\over 45 } -{M^2T^2\over 12}
-{4\pi T^3\over a}e^{-{4\pi aT\sqrt{1+{M^2\over4\pi^2T^2}} }}-{\zeta_R(3)T\over 4\pi a^3}
\label{end_14}
\end{equation}
valid for $M\ll a^{-1}$ and where $\zeta_R(3)=1.20206$,

  \begin{equation}
\bar{P}_C={\pi^{2}T^4\over 45 } -{M^2T^2\over 12}
-{4\pi T^3\over a}e^{-{4\pi aT\sqrt{1+{M^2\over4\pi^2T^2}} }}-{TM^2\over 2\pi a}e^{-2Ma}
\label{end_14b}
\end{equation}
valid for $M\gg a^{-1}$.

 \underline{ Large mass}
  \begin{equation}
\bar{P}_C=-{M^{3/2}\over 4\pi^{3/2}}\left({M\over a^{3/2}}e^{-2Ma}-2^{3/2}T^{5/2}e^{-{M\over T}}\right).
\label{end_15}
\end{equation}

Notice that, while the Casimir pressure without Lorentz violation is attractive in the small plate distance and large mass asymptotic limits and repulsive in the high temperature limit, the Lorentz violating correction for $\epsilon = 2$ weakens the Casimir pressure in all three asymptotic limits and for $u^\mu$ parallel and perpendicular to the plates. The $\epsilon = 3 $ Lorentz violating correction increases the Casimir pressure in all three limits and for both directions of $u^\mu$, but with one exception: the large mass limit when $u^\mu$ is perpendicular to the plates, in which case the attractive pressure is weakened by the Lorentz violating correction. I find that what I observe for $\epsilon =2$ is true for all even values of $\epsilon$, and what I observe for $\epsilon = 3$ is true for all odd values of $\epsilon$.

Finally, a brief comment on the Casimir pressure under mixed boundary conditions. In this case, the small plate distance and large mass limits have reversed signs when compared to the case of Dirichlet boundary conditions. The Casimir pressure without Lorentz violation is repulsive in those two asymptotic limits, and the Lorentz violating corrections are mostly attractive. With the exception of minor differences in the numerical coefficients in the short plate distance limit, the Lorentz violating correction to the Casimir pressure is the opposite of the Lorentz violating correction under Dirichlet boundary conditions. The high temperature limit of the Casimir pressure and its Lorentz violating corrections behave qualitativley as they do under Dirichlet boundary conditions, with some exceptions on the numerical coefficients.


\begin{thebibliography}{99}

\bibitem{Casimir:1948dh}
  H.~B.~G.~Casimir,
  Indag.\ Math.\  {\bf 10}, 261 (1948)
  [Kon.\ Ned.\ Akad.\ Wetensch.\ Proc.\  {\bf 51}, 793 (1948)]
  [Front.\ Phys.\  {\bf 65}, 342 (1987)]
  [Kon.\ Ned.\ Akad.\ Wetensch.\ Proc.\  {\bf 100N3-4}, 61 (1997)].

\bibitem{Sparnaay:1958wg}
  M.~J.~Sparnaay,
  Physica {\bf 24}, 751 (1958).

\bibitem{Bordag:2001qi}
  M.~Bordag, U.~Mohideen, and V.~M.~Mostepanenko,
  Phys.\ Rep.\  {\bf 353}, 1 (2001).

\bibitem{Bordag:2009zz} 
  M.~Bordag, G.~L.~Klimchitskaya, U.~Mohideen, and V.~M.~Mostepanenko,
  {\it Advances in the Casimir effect},
  International Series of Monographs on Physics, Vol. {\bf 145} (Oxford University, New York, 2009), p. 1.
 
\bibitem{Boyer:1974}
  T.~H.~Boyer,
  Phys.\ Rev.\  A {\bf 9}, 2078 (1974).

\bibitem{Boyer:1968uf}
  T.~H.~Boyer,
{\it   Phys.\ Rev.}\  {\bf 174}, 1764 (1968).

\bibitem{Ferrari:2010dj}
A.~Ferrari, H.~Girotti, M.~Gomes, A.~Petrov and A.~da Silva,
Mod. Phys. Lett. A \textbf{28}, 1350052 (2013).

\bibitem{Ulion:2015kjx}
I.~Morales Ulion, E.~Bezerra de Mello and A.~Petrov,
Int. J. Mod. Phys. A \textbf{30}, no.36, 1550220 (2015).

\bibitem{Alfaro:1999wd}
J.~Alfaro, H.~A.~Morales-Tecotl and L.~F.~Urrutia,
Phys. Rev. Lett. \textbf{84}, 2318-2321 (2000).

\bibitem{Alfaro:2001rb}
J.~Alfaro, H.~A.~Morales-Tecotl and L.~F.~Urrutia,
Phys. Rev. D \textbf{65}, 103509 (2002).

\bibitem{Kostelecky:2002ca}
V.~Kostelecky, R.~Lehnert and M.~J.~Perry,
Phys. Rev. D \textbf{68}, 123511 (2003).

\bibitem{Anchordoqui:2003ij}
L.~Anchordoqui and H.~Goldberg,
Phys. Rev. D \textbf{68}, 083513 (2003).

\bibitem{Bertolami:1997iy}
O.~Bertolami,
Class. Quant. Grav. \textbf{14}, 2785-2791 (1997).

\bibitem{Kostelecky:1988zi}
V.~Kostelecky and S.~Samuel,
Phys. Rev. D \textbf{39}, 683 (1989).

\bibitem{Cruz:2017kfo} 
  M.~B.~Cruz, E.~R.~Bezerra de Mello and A.~Y.~Petrov,
  Phys.\ Rev.\ D {\bf 96}, no. 4, 045019 (2017).

\bibitem{ADantas:2023wxd}
R.~A.~Dantas, H.~F.~S.~Mota and E.~R.~Bezerra de Mello,
Universe \textbf{9}, no.5, 241 (2023).

\bibitem{Cruz:2018bqt}
M.~Cruz, E.~Bezerra De Mello and A.~Y.~Petrov,
Mod. Phys. Lett. A \textbf{33}, no.20, 1850115 (2018).

\bibitem{daSilva:2019iwn}
D.~R.~da Silva, M.~B.~Cruz and E.~R.~Bezerra de Mello,
{\it  Int. J. Mod. Phys. A} \textbf{34}, no.20, 1950107 (2019).

\bibitem{Frank:2006ww}
M.~Frank and I.~Turan,
Phys. Rev. D \textbf{74}, 033016 (2006).

\bibitem{Kharlanov:2009pv}
O.~Kharlanov and V.~Zhukovsky,
Phys. Rev. D \textbf{81}, 025015 (2010).

\bibitem{Martin-Ruiz:2016lyy}
A.~Martin-Ruiz and C.~Escobar,
Phys. Rev. D \textbf{95}, no.3, 036011 (2017).

\bibitem{Cruz:2020zkc}
M.~B.~Cruz, E.~R.~Bezerra de Mello and H.~F.~Santana Mota,
Phys. Rev. D \textbf{102}, no.4, 045006 (2020)

\bibitem{Erdas:2020ilo}
A.~Erdas,
Int. J. Mod. Phys. A \textbf{35}, no.31, 2050209 (2020).

\bibitem{Erdas:2021xvv}
A.~Erdas,
Int. J. Mod. Phys. A \textbf{36}, no.20, 20 (2021).

\bibitem{Fierz:1960zq}
M.~Fierz,
Helv. Phys. Acta \textbf{33}, 855-858 (1960).

\bibitem{Mehra:1967wf}
J.~Mehra,
Physica \textbf{37}, 145-152 (1967).

\bibitem{CougoPinto:1998jg} 
  M.~V.~Cougo-Pinto, C.~Farina, M.~R.~Negrao, and A.~Tort,
  arXiv:hep-th/9810033.

\bibitem{Erdas:2013jga} 
  A.~Erdas and K.~P.~Seltzer,
  Phys.\ Rev.\ D {\bf 88}, 105007 (2013).

\bibitem{Erdas:2013dha} 
  A.~Erdas and K.~P.~Seltzer,
  Int.\ J.\ Mod.\ Phys.\ A {\bf 29}, 1450091 (2014).

\bibitem{Aleixo:2021cfy}
G.~Aleixo and H.~F.~S.~Mota,
Phys. Rev. D \textbf{104}, no.4, 045012 (2021).

\bibitem{Hawking:1976ja} S.~W.~Hawking,
{\it    Commun.\ Math.\ Phys.\ } {\bf 55}, 133 (1977).

\bibitem{Elizalde:1988rh}
  E.~Elizalde and A.~Romeo,
  J.\ Math.\ Phys.\  {\bf 30}, 1133 (1989)
  [Erratum-ibid.\  {\bf 31}, 771 (1990)].
 
\bibitem{Elizalde:2007du}
  E.~Elizalde,
  J.\ Phys.\ A  {\bf 41}, 304040 (2008).

\bibitem{Farias:2011aa}
C.~F.~Farias, M.~Gomes, J.~R.~Nascimento, A.~Y.~Petrov and A.~J.~da Silva,
{\it  Phys. Rev. D} \textbf{85}, 127701 (2012).

\bibitem{Erdas:2025etj}
A.~Erdas,
Int. J. Mod. Phys. A \textbf{40}, no.15, 2550040 (2025).

\bibitem{Erdas:2025gbv}
A.~Erdas,
Int. J. Mod. Phys. A \textbf{40}, no.19, 2550061 (2025).
\end{thebibliography}
\end{document}